\documentclass{article}

\usepackage{arxiv}

\usepackage[utf8]{inputenc} 
\usepackage[T1]{fontenc}    
\usepackage{url}            
\usepackage{booktabs}       
\usepackage{amsfonts}       
\usepackage{nicefrac}       
\usepackage{microtype}      
\usepackage{graphicx}
\usepackage{doi}
\usepackage[all]{xy}
\usepackage{setspace}
\usepackage{xcolor}
\usepackage{tikz}
\usetikzlibrary{cd}
\usepackage{amsmath,amssymb,amsfonts,latexsym,float,graphics,epsfig}
\usepackage[colorinlistoftodos,textsize=tiny,obeyFinal]{todonotes}
\usepackage{epstopdf}
\usepackage{comment}

\newcommand{\subf}[2]{%
  {\small\begin{tabular}[t]{@{}c@{}}
  #1\\#2
  \end{tabular}}%
}

\begin{document}
\title{Approaches to conservative Smoothed Particle Hydrodynamics with entropy}

\author{Michal Pavelka\\
 Mathematical Institute, Faculty of Mathematics and Physics, Charles University,\\ 
 Sokolovsk\'{a} 83, 18675 Prague, Czech Republic\\
Corresponding author: pavelka@karlin.mff.cuni.cz\\
\And Václav Klika\\
Department of Mathematics, FNSPE, Czech Technical University in Prague\\
Trojanova 13, Prague 2, 120 00, Czech Republic\\
\And
and Ondřej Kincl\\
 Mathematical Institute, Faculty of Mathematics and Physics, Charles University,\\ 
 Sokolovsk\'{a} 83, 18675 Prague, Czech Republic\\
Department of Civil, Environmental and Mechanical Engineering, University of Trento,\\
Via Mesiano 77, Trento, 38123, Italy
}


\newcommand{\shorttitle}{Entropic SPH}

\hypersetup{
pdftitle={Entropic SPH},
pdfsubject={},
pdfauthor={M. Pavelka, V. Klika, and O. Kincl},
pdfkeywords={smoothed particle hydrodynamics, entropy, Hamiltonian mechanics}
}

\maketitle

\newcommand{\qq}{\mathbf{q}}
\newcommand{\rr}{\mathbf{r}}
\newcommand{\pp}{\mathbf{p}}
\newcommand{\ww}{\mathbf{w}}
\newcommand{\mm}{\mathbf{m}}
\newcommand{\xx}{\mathbf{x}}
\newcommand{\ee}{\mathbf{e}}
\newcommand{\XX}{\mathbf{X}}
\newcommand{\FF}{\mathbf{F}}
\newcommand{\MM}{\mathbf{M}}
\newcommand{\ppi}{\boldsymbol{\pi}}
\newcommand{\vv}{\mathbf{v}}
\newcommand{\uu}{\mathbf{u}}
\newcommand{\chibar}{\bar{\chi}}


\doublespace

\begin{abstract}
Smoothed particle hydrodynamics (SPH) is typically used for barotropic fluids, where the pressure depends only on the local mass density. Here, we show how to incorporate the entropy into the SPH, so that the pressure can also depend on the temperature, while keeping the growth of the total entropy, conservation of the total energy, and symplecticity of the reversible part of the SPH equations. The SPH system of ordinary differential equations with entropy is derived by means of the Poisson reduction and the Lagrange $\rightarrow$ Euler transformation. We present several approaches towards SPH with entropy, which are then illustrated on systems with discontinuities, on adiabatic and nonadiabatic expansion, and on the Rayleigh-Bénard convection without the Boussinesq approximation. Finally, we show how to model hyperbolic heat conduction within the SPH, extending the SPH variables with not only entropy but also a heat-flux-related vector field.
\end{abstract}

\tableofcontents

\section{Introduction}
Smoothed Particle Hydrodynamics (SPH) is a particle-based discretization of the hydrodynamic partial differential equations \cite{sph,lucy}. Each particle is typically described by its position and velocity, and the particles interact in a way that mimics the hydrodynamic behavior. Although simulation can be carried out in a purely reversible way (being even symplectic and globally reversible \cite{sph-reversible}), viscous effects reduce the kinetic energy of the particles and the total energy is often not conserved by SPH schemes. The addition of entropy to each particle should restore total energy conservation \cite{violeau,monaghan-review}. In this paper, we show several approaches towards SPH with entropy of each particle while keeping the geometric structure of the reversible part of the SPH equations (symplecticity).

Cleary and Monaghan \cite{cleary1999} add the internal energy to each particle, the evolution of which compensates the dissipation by Fourier heat conduction so that the total energy is conserved. However, the integration scheme is not symplectic. In \cite{antuono2010}, the energy equation contains a term that violates global energy conservation (the second term on the right-hand side of the energy equation in Eqs. (12) therein). In \cite{antuono2015}, the $\delta$-SPH scheme does not contain energy or entropy of individual particles. The compressional and kinetic energies are not conserved (nor their sum is), so the total energy is not conserved in that scheme. In general, the total energy can not be conserved if the SPH particles possess no entropy, internal energy, or temperature.

However, even when SPH particles are equipped with entropy or internal energy \cite{elleropre}, another source of irreversibility prevails in the nonsymplecticity of the underlying numerical scheme. Consequently, the solution of the SPH numerical scheme is irreversible even if dissipative terms (viscosity and heat conductivity) are omitted. 

In this paper, we combine symplectic integrators for SPH particles equipped with entropy and conservative dissipative dynamics. Numerical illustrations contain adiabatic and nonadiabatic expansion of ideal gas, thermal convection of stiffened gas within a container heated from below, and hyperbolic heat conduction. The global energy errors are typically of order $10^{-5}\%$ in the nondissipative case and $10^{-3}\%$ in the dissipative case (using only first-order discretization of the dissipative terms). 

Section \ref{sec.Ham} recalls the usual reversible (and symplectic) formulation of SPH without entropy, but it shows how to derive the evolution equations by a reduction of Poisson brackets. In particular, the Lagrangian continuum mechanics is discretized to Lagrangian particles (interpreted as the SPH particles). Section \ref{sec.entropy} shows five approaches to the entropy density in SPH, based on four definitions of particle volumes (mass-based, entropy-based, direct, and implicit), as well as a mixed definition. Section \ref{sec.ham.ent} contains the reversible (Hamiltonian) part of the five approaches to SPH with entropy, mimicking the derivation of the Poisson bracket from Section \ref{sec.Ham}. Finally, in Section \ref{sec.dis} we add dissipative (non-Hamiltonian) terms for viscosity, Fourier heat conduction (using gradient dynamics), and hyperbolic heat conduction.

The problematic of discontinuities (shocks) in a compressible fluid is not covered in this paper. The most common approach to shocks in SPH is via artificial viscosity \cite{cullen2010inviscid}. When it comes to Riemann solvers, they are difficult to combine with SPH. Nonetheless, this direction was investigated by Inutsuka \cite{inutsuka2002reformulation}. 

\section{SPH without entropy as a Hamiltonian system}\label{sec.Ham}
Both the standard SPH and the SPH with entropy are Hamiltonian systems equipped with dissipative terms (viscosity, heat conduction, etc.). Before approaching the derivation of the reversible (Hamiltonian) part of the entropic SPH, let us first recall the Hamiltonian formulation of continuum mechanics, which serves as the starting point of the derivation of the usual SPH without entropy.

\subsection{Hamiltonian continuum mechanics}\label{sec.Lag}
Hamiltonian continuum mechanics on the Lagrangian (reference) manifold can be derived by means of the principle of stationary action. The Lagrangian is a function of the Lagrange$\rightarrow$Euler mapping $\xx(t,\XX)$ and its time-derivative, where $t$ is the time, $\XX$ the Lagrangian coordinates of a point, and $\xx$ its Eulerian coordinate (current configuration) \cite{Goldstein}. The principle of stationary action then gives the Lagrange-Euler equation, which can be rewritten (by the Legendre transformation) to the Hamiltonian setting with the fields of position and momentum density, $x^i(\XX)$ and $M_i(\XX)$, as the state variables. These state variables are equipped with the canonical Poisson bracket
\begin{equation}\label{eq.PB.Lag}
    \{F,G\}^{(\mathrm{Lagrangian})} = \int d\XX \left(F_{x^i}G_{M_i} - G_{x^i}F_{M_i}\right),
\end{equation}
and Hamiltonian evolution of any functional of the state variables, $F(\xx,\MM)$, is given by the Poisson bracket of the functional and the energy $E(\xx,\MM)$,
\begin{equation}
\dot{F} = \{F,E\}.
\end{equation}
In particular, if $F$ is chosen as the state variables, we obtain the Hamiltonian evolution equations of the positions and momenta, 
\begin{subequations}\label{eq.evo.Lag}
    \begin{align}
        \dot{\xx} &= E_{\MM}\\
        \dot{\MM} &= -E_{\xx},
    \end{align}
where the subscripts denote derivatives (more precisely, the functional derivatives \cite{pkg}). 
\end{subequations}
Equations \eqref{eq.evo.Lag} represent the canonical Hamiltonian equations for the Lagrangian continuum mechanics, which can be used as dynamics of elastic bodies \cite{landau7,LagEul}.

However, to close the equations, we have to prescribe the energy $E(\xx,\MM)$. If we know the internal energy density per the Eulerian volume $\epsilon(\rho)$, the Lagrangian energy is the sum of the kinetic energy and the internal energy transformed to the Lagrangian manifold,
\begin{equation}\label{eq.E.Lag}
    E^{Lagrangian} = \int d\XX \left(\frac{\MM^2}{2\rho_0} + \det \frac{\partial \xx}{\partial \XX} \epsilon(\rho(\xx(\XX)))\right).
\end{equation}
Here, $\rho$ is the Eulerian mass density, which is related to the Lagrangian density $\rho_0(\XX)$ (mass density per the Lagrangian volume) through
 \begin{equation}
     \rho(\xx) = \rho_0(\XX(\xx)) \det \frac{\partial \XX}{\partial \xx}.
 \end{equation}
Without going to more details on the Hamiltonian continuum mechanics on the Lagrangian manifold, let us go to the Hamiltonian formulation of the standard SPH method.

\subsection{Hamiltonian form of the standard weakly compressible SPH}
The standard SPH \cite{sph,violeau} expresses evolution of positions and velocities of the smoothed particles, and the reversible part of the evolution (disregarding viscosity) is also a Hamiltonian system.
To write down the classical SPH formulation, we need two approximations: 
\begin{enumerate}
    \item smoothed state variables (positions and momenta of the particles)
    \item differential operators (appearing in the evolution equations like $\mathrm{div}$ or $\mathrm{grad}$)
\end{enumerate}
Different operator discretization may yield different results as clearly observed in the discussion of density evolution \cite[Chap 5.3.1]{violeau}. 
Here, we propose quite a different approach, employing the least unnecessary knowledge for deriving discrete operators, thus reducing ambiguity in the formulation of SPH. 

\begin{figure}[ht!]
\begin{center}
\includegraphics[scale=0.5]{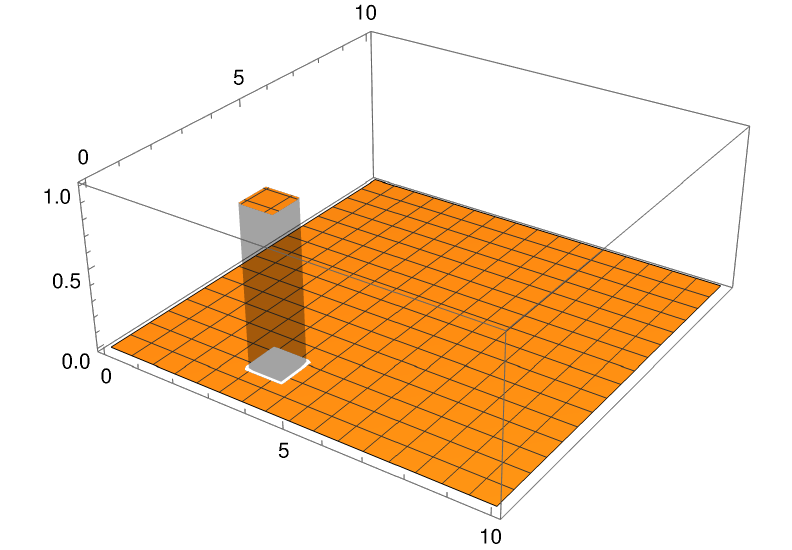}
\caption{\label{fig.chi}An example of partitioning of the Lagrangian manifold into Lagrangian particles, which then become the SPH particles.}
\end{center}
\end{figure}

Instead of discretizing the differential operators, we derive SPH by Poisson reduction from the Hamiltonian form of Lagrangian continuum mechanics, recalled in Section \ref{sec.Lag}. To this end, we first need a mapping projecting the continuum state variables $\xx$ and $\MM$ to discrete SPH positions and momenta.
Let $\Omega_0$ be the Lagrangian manifold, which can be split into mutually disjoint cells (for instance a grid in 2D, see Fig. \ref{fig.chi}), $\Omega_0 = \bigcup_\alpha \Omega_{0 \alpha}$. Each of $\Omega_{0 \alpha}$ represents a Lagrangian particle, that will become an SPH particle. The volume of Lagrangian cell with index $\alpha$ is 
\begin{subequations}
\begin{equation}
    V_{0\alpha} = \int_{\Omega_{0\alpha}} d\XX,
\end{equation}
and we will also need the normalized characteristic functions of the Lagrangian particles 
\begin{equation}
    \chibar_\alpha(\XX) = 
    \begin{cases}
        \frac{1}{V_{0\alpha}} & \text{if} \quad \XX \in \Omega_{0\alpha}\\
        0 & \text{otherwise}
    \end{cases}.
\end{equation}
\end{subequations} 

The mapping 
\begin{subequations}\label{eq.projection}
    \begin{align}
        \xx_{\alpha} &= \int d\XX \chibar_\alpha(\XX) \xx(\XX)\\
        \MM_{\alpha} &= \int d\XX V_{0\alpha} \chibar_\alpha(\XX) \MM(\XX)
    \end{align}
then defines the position and momentum of the SPH particle $\alpha$, which play the role of the SPH state variables. Later, in Section \ref{sec.entropy}, we will also add entropy to the state variables, but first we show how to derive the standard SPH by reduction from the Lagrangian continuum mechanics.
\end{subequations}

In order to reduce the continuum Poisson bracket \eqref{eq.PB.Lag} to a Poisson bracket for the SPH state variables ($\xx_\alpha$ and $\MM_\alpha$), we plug functionals dependent on the SPH variables into the Poisson bracket. 
Because the Poisson bracket contains derivatives of the functionals, we need to calculate derivatives of the SPH state variables with respect to the continuum state variables,
\begin{equation}
    \frac{\delta x^i_\alpha}{\delta x^j(\XX)} = \delta^i_j \chibar_\alpha(\XX)
    \quad\text{and}\quad
    \frac{\delta M_{\alpha i}}{\delta M_j(\XX)} = \delta^j_i V_{0\alpha}\chibar_\alpha(\XX)
\end{equation}
while the remaining derivatives are zero. Derivatives of an arbitrary functional of the SPH state variables, $F(\xx_\alpha, \MM_\alpha)$, are then
\begin{equation}
    \frac{\delta F}{\delta x^i(\XX)} = \sum_\alpha \frac{\partial F}{\partial x^i_\alpha}\chibar_\alpha(\XX)
    \quad\text{and}\quad
    \frac{\delta F}{\delta M_i(\XX)} = \sum_\beta \frac{\partial F}{\partial M_{\beta i}}V_{0\beta}\chibar_\beta(\XX).
\end{equation}
Plugging two arbitrary functionals dependent on the discrete state variables to Poisson bracket \eqref{eq.PB.Lag} leads to
\begin{align} \label{eq.PB.SPH}
    \{F,G\}^{(\mathrm{SPH})} &= \sum_\alpha \sum_\beta \left(F_{x^i_{\alpha}} G_{M_{\beta i}}-G_{x^i_{\alpha}} F_{M_{\beta i}} \right) \int d\XX V_{0\beta}\chibar_\alpha(\XX)\chibar_\beta(\XX)\nonumber\\
    &= \sum_\alpha \left(F_{x^i_{\alpha}} G_{M_{\alpha i}}-G_{x^i_{\alpha}} F_{M_{\alpha i}} \right),
\end{align}
which is the SPH Poisson bracket expressing kinematics of the SPH state variables. This Poisson bracket can be seen as the Hamiltonian counterpart of the Lagrangian form of SPH \cite{violeau}.

The reversible Hamiltonian evolution equations implied by this bracket are
\begin{subequations}\label{eq.SPH}
    \begin{align}
        \dot{\xx}_\alpha &= E_{\MM_\alpha}\\
        \dot{\MM}_\alpha &= -E_{\xx_\alpha},
    \end{align}
which are the Hamilton canonical equations for the SPH state variables. 
\end{subequations}
However, to close the equations, we need to express the energy in terms of the SPH state variables.

\subsection{Energy functional in the standard SPH}
To find the dependence of the energy on the SPH state variables, we need to approximate the exact energy of the continuous system \eqref{eq.E.Lag}, which depends also on the full Lagrange$\rightarrow$Euler mapping $\xx(\XX)$. Knowledge of discrete particle positions is not enough for the precise reconstruction of the exact continuum energy, and hence we turn to the standard SPH smoothing using the weighing functions $W$:
\begin{equation} \label{eq.DefOfDensVar}
 \overline{A}_\alpha = \sum_\beta A_\beta W(|\xx_\alpha-\xx_\beta|),
\end{equation}
which is an approximate form of mollification. It can be shown \cite{violeau} that if the weighing function satisfies certain normalization property, it provides an approximation precise up to the second-order. The normalization requirement can be translated into a condition that $W$ is a homogeneous function of the distance of a degree $-n$, that is $W(\alpha|\xx|)=\alpha^{-n} W(|\xx|)$, where $n$ denotes the spatial dimension. From this homogeneous property, one can also see that any scaling of the local volume (as in bulk expansion) is exactly captured by the weighing function $W$. Finally, as the dimension of $W$ is $1/m^n$, we may define any density variable discretization in Eulerian frame $\overline{A}_{\alpha}$ from its Lagrangian counterpart $A_{\alpha}$ via the above expression \eqref{eq.DefOfDensVar}.
We shall use such smoothing to define the SPH mass density $\rho_\alpha$. In particular, $\rho_\alpha$ is the Eulerian mass density of particle $\alpha$, $m_\alpha = \int d\XX \rho_{0}(\XX) \chibar_\alpha(\XX)$ is the mass of the Lagrangian particle and hence
\begin{equation}\label{eq.rhoa}
    \rho_\alpha =  \sum_\beta m_\beta W_{\alpha\beta}, 
\end{equation}
where $W_{\alpha\beta}\stackrel{def}{=}W(|\xx_\alpha-\xx_\beta|)$.

Now we can close the Hamilton canonical equations \eqref{eq.SPH} by supplying an energy functional that depends only on the SPH state variables. This is actually the step that depends on material properties, since the Poisson bracket \eqref{eq.PB.SPH} is already fixed by the SPH state variables. The continuum energy \eqref{eq.E.Lag} can be approximated by 
 \begin{align}\label{eq.SPH.E}
   E^{SPH} &= \sum_\alpha V_{0\alpha}\frac{\MM^2_\alpha}{2\rho_{0\alpha}} +\sum_\alpha V_{0\alpha}\frac{\rho_{0\alpha}}{\rho_\alpha} \epsilon(\rho_\alpha)\nonumber\\
     &= \sum_\alpha \frac{\MM^2_\alpha}{2m_{\alpha}}+\sum_\alpha V_{\alpha} \epsilon(\rho_\alpha),
\end{align}
where $\rho_{0\alpha} = m_\alpha/V_{0\alpha}$ is its density in the Lagrangian space. This approximate energy functional can also be obtained by means of the principle of maximum entropy (MaxEnt), which is shown in Appendix \ref{sec.MaxEnt}.
We keep the Eulerian volume $V_{\alpha}$ unspecified (unrelated to Eulerian density, for example), as it is a key step subjected to further in Section \ref{sec.entropy}, where five possible definitions are shown.

\subsection{Standard SPH as a Hamiltonian system}
The purpose of this Section is to collect the above results on the SPH Poisson bracket and SPH energy and to show that they indeed lead to the standard formulation of SPH. 

When energy \eqref{eq.SPH.E} is plugged into the Hamilton canonical equations \eqref{eq.SPH}, we need to take its derivatives with respect to the SPH state variables. In particular, derivative of $\rho_\alpha$ (defined in Equation \eqref{eq.rhoa}) with respect to the positions reads
\begin{align}
    \frac{\partial \rho_\beta}{\partial x^i_\alpha} &=  \sum_{\gamma} m_\gamma W'(|\xx_\beta-\xx_\gamma|) \frac{\partial \sqrt{(x^j_\beta-x^j_\gamma)(x^j_\beta-x^j_\gamma)}}{\partial x^i_\alpha}\nonumber\\
    &= \sum_{\gamma} m_\gamma W'(|\xx_\beta-\xx_\gamma|) \frac{x^j_\beta-x^j_\gamma}{|\xx_\beta-\xx_\gamma|}\delta^j_i\left(\delta_{\beta\alpha}-\delta_{\gamma\alpha}\right)\nonumber\\
    &=  \sum_{\gamma} m_\gamma W'_{\beta\gamma}e_{i\beta\gamma}\left(\delta_{\beta\alpha}-\delta_{\gamma\alpha}\right),
\end{align}
where $W'_{\beta\gamma} = W'(|\xx_\beta-\xx_\gamma|)$ is the derivative of the SPH kernel $W$ evaluated on the distance between particles $\beta$ and $\gamma$ and where $\ee_{\beta\gamma} = \frac{\xx_\beta-\xx_\gamma}{|\xx_\beta-\xx_\gamma|}$ is the unit vector\footnote{Actually, it is the covector expressing gradient of the distance between the particles while the vector is dual to the covector. But in Euclidean space we do not need to distinguish them \cite{fecko}.} pointing from $\xx_\gamma$ to $\xx_\beta$.

Now, we are finally in position to write down the resulting ordinary differential equations expressing the Hamiltonian evolution of the SPH state variables,
\begin{subequations}\label{eq.SPH.symplectic}
    \begin{align}
        \dot{x}^i_\alpha &= \frac{M^i_\alpha}{m_\alpha}\\
      \dot{M}_{\alpha i} &= -\sum_\beta\left(-\frac{m_\beta}{\rho^2_\beta} \epsilon + \frac{m_\beta}{\rho_\beta}\frac{\partial \epsilon}{\partial \rho_\beta}\right)  \sum_\gamma m_\gamma W'_{\beta\gamma}e_{i\beta\gamma}(\delta_{\beta\alpha}-\delta_{\gamma\alpha})\nonumber\\
          &= -\sum_\beta\left(\frac{m_\alpha m_\beta}{\rho^2_\alpha}p_\alpha + \frac{m_\alpha m_\beta}{\rho^2_\beta}p_\beta\right) W'_{\alpha\beta}e_{i\alpha \beta},
    \end{align}
\end{subequations}
where the pressure of particle $\alpha$ is defined by the usual barotropic relation
\begin{equation}\label{eq.p}
    p_\alpha = -\epsilon + \rho_\alpha \frac{\partial \epsilon}{\partial \rho_\alpha},
\end{equation}
see \cite{callen,pkg}.
When the dependence of the internal energy on the mass density is prescribed, which can be integrated from an equation of state, the system of equations is completely specified. See \cite{violeau} for some choices of internal energy typical in SPH. 

Equations \eqref{eq.SPH.symplectic} represent a symplectic SPH discretization of fluid mechanics where the energy depends on the mass density \cite{violeau}. This system of equations forms a Hamiltonian system because it is constructed from a Poisson bracket and energy. Combined with a symplectic time integrator, such as the Verlet scheme, the numerical solution has energy error bounded uniformly with respect to the simulation time $t$, and the numerical schemes can even be made time reversible by using fixed-point arithmetic \cite{sph-reversible}. 

However, it is customary in the SPH literature \cite{sph} to use the discretized continuity equation instead of the closed expression \eqref{eq.rhoa}. Then, Equations \eqref{eq.SPH.symplectic} can be generalized, using the definition of $\rho_a$, to another Hamiltonian system,
\begin{subequations}\label{eq.SPH.standard}
    \begin{align}
      \dot{x}^i_\alpha &= \frac{M^i_\alpha}{m_\alpha}\\
      \dot{\rho}_\alpha &= \sum_\beta m_\beta W'_{\alpha\beta}\ee_{\alpha\beta}\left(\frac{\MM_\alpha}{m_\alpha} - \frac{\MM_\beta}{m_\beta}\right)\\
      \dot{M}_{\alpha i} &= -\sum_\beta\left(\frac{m_\alpha m_\beta}{\rho^2_\alpha}p_\alpha + \frac{m_\alpha m_\beta}{\rho^2_\beta}p_\beta\right) W'_{\alpha\beta}e_{i\alpha \beta},
    \end{align}
where the pressure is again given by Equation \eqref{eq.p}. Equations \eqref{eq.SPH.standard} are the standard SPH equations \cite{violeau}.
\end{subequations}
This system of equations is Hamiltonian because it is generated by a Poisson bracket
\begin{align}\label{eq.PB.SPH.rhoa}
\{F,G\}^{(\mathrm{SPH},\rho_\alpha)}&= \{F,G\}^{(\mathrm{SPH})} \nonumber\\
    &+ \sum_\alpha\sum_\beta m_\beta W'_{\alpha\beta}e_{i\alpha\beta}\left(F_{\rho_\alpha}(G_{M_{\alpha i}}-G_{M_{\beta i}})
    -G_{\rho_\alpha}(F_{M_{\alpha i}}-F_{M_{\beta i}})\right),
\end{align}
which can be obtained by projection of the continuum Poisson bracket \eqref{eq.PB.Lag} to the state variables $\xx_\alpha$, $\MM_\alpha$, and $\rho_\alpha$, see Appendix \ref{sec.PB} for details (here we dropped $s_{\alpha}$ from state variables considered in Appendix \ref{sec.PB}). 

Poisson bracket \eqref{eq.PB.SPH.rhoa} is different from the bracket \eqref{eq.PB.SPH} and leads to Hamiltonian but non-symplectic mechanics \cite{marsden-ratiu}. 
From a numerical perspective, it is advantageous to keep the symplecticity (non-degenerate Poisson bracket) because, for instance, the energy error is then capped by a constant decreasing with the time step when a symplectic integrator is used \cite{hairer,leimkuehler}. Therefore, we make the dynamics symplectic by directly evaluating the density from its definition \eqref{eq.rhoa} at each time step while keeping the evolution equations for $\xx_\alpha$ and $\MM_\alpha$. This results, for example, in a globally reversible scheme for SPH \cite{sph-reversible}. On the other hand, this direct evaluation of the density within the boundaries may cause other numerical artifacts \cite{antuono2015} if not properly treated \cite{sph-reversible}.

In summary, the standard SPH equations \eqref{eq.SPH.standard} can be derived from Lagrangian continuum mechanics by reducing the continuum Poisson bracket \eqref{eq.PB.Lag} to the SPH Poisson bracket \eqref{eq.PB.SPH.rhoa}. The reduction relies on the projection from the continuous fields of positions and momenta to the SPH positions and momenta \eqref{eq.projection}, where the Lagrangian manifold is partitioned into discrete particles. Then, the SPH energy \eqref{eq.SPH.E} is found as an approximation of the continuum energy \eqref{eq.E.Lag}, and this approximation can be seen as an application of the principle of the MaxEnt principle (Sect. \ref{sec.MaxEnt}). When a concrete dependence of the internal energy on the mass density is chosen (a material is chosen), the SPH equations \eqref{eq.SPH.standard} can be solved as long as one chooses an appropriate interpretation of the particle volume. 

However, what if the energy also depends on the entropy of the material, which means that the fluid is not barotropic? Then, the formulation of SPH has to be enriched to include also the entropy of the SPH particles, which is the purpose of the following Section.

\section{Approximations of particle volumes, mass density, and entropy density}\label{sec.entropy}
The usual formulation of SPH \eqref{eq.SPH.standard} does not involve the entropy of the particles, which in particular means that the pressure depends only on the density, and thus the fluid is barotropic. In reality, however, the pressure depends on both the density and the entropy (or temperature) of the fluid, and this dependence is captured by more precise models (see, for instance, the ideal gas or the stiffened gas in Appendix \ref{sec.thermo}).  
We include entropy in SPH as a variable of the density of the volume. However, this highlights the issue with various possible definitions of SPH particle volumes, which we discuss now. For example, even the very definition of SPH energy \eqref{eq.SPH.E} depends on such choice as is immediate from its extension to the case with entropy. 
For nonbarotropic fluids, the energy density depends on the entropy density, and hence the expression for energy reads as 
 \begin{align}\label{eq.SPH.ES}
     E^{SPH} &= \sum_\alpha V_{0\alpha}\frac{\MM^2_\alpha}{2\rho_{0\alpha}} + \sum_\alpha V_{0\alpha}\frac{V_\alpha}{V_{0\alpha}}\epsilon(\rho_\alpha, s_\alpha)\nonumber\\
     &=   \sum_\alpha \frac{\MM^2_\alpha}{2 m_\alpha} +\sum_\alpha V_\alpha \epsilon(\rho_\alpha, s_\alpha),
 \end{align}
 where  $s_\alpha$ is the Eulerian entropy density of particle $\alpha$.

 Consider a volume density state variable as is standard continuum mechanics which we discretize to the SPH particles. We denote its Lagrangian variant by $Z_{\alpha}(X)$ (typically prescribed as an initial condition), and then its Eulerian counterpart $\zeta_{\alpha}(x)$ is then related via
 \begin{equation}
   \label{eq:LagEulStateVars}
   \zeta_{\alpha}(x) = \frac{V_{0\alpha}}{V_{\alpha}} Z_{\alpha}(X).
 \end{equation}
The Lagrangian particle volume follows directly from the SPH discretization but the Eulerian volume can be estimated in several distinct ways, for example, from the update of particle positions $x_{\alpha}$ or from the evolution of any volume density state variable. This gives two possibilities of particle volume evolution in the classical SPH formulations (the second one stemming from mass density) but the introduction of entropy density introduces another choice. They all represent admissible approximation of particle volume but differ in practice.

Hence, naturally, we start with a discussion of approximation errors.

\subsection{SPH discretization error estimates}

The fundamental idea behind SPH discretization lies in approximative formulas for the Dirac delta function and convolution. In particular, if we take a classical integrable and normalized function, that is, $\int_{\mathbb{R}}W=1$, and define a rescaled function as $W_{h}(x)=\frac{1}{h} W(x/h)$, then $\int_{\mathbb{R}} W_{h}=1$ and $\lim_{h\to 0} W_{h}=\delta(x)$. As $f\star \delta=f$, we get $  f\approx f\star W_h$. In particular, we have
\begin{equation*}
f(x_\alpha) \approx \int_{\mathbb{R}^n} f(y) W_h(x-y) \approx \sum_\beta V_\beta f(x_\beta) W_h(x_\alpha-x_\beta).
\end{equation*}
With particular choices of the approximated function $f$, we obtain different approximation formulas for volume and different corresponding errors.

For the Wendland kernel, one can show that the relative error in 1D is proportional to the gradient of a logarithm at that point \cite[Eq. (5.115)]{violeau}
\begin{equation}\label{eq:Error}
  \frac{f(x)-f(x_\alpha)}{f(x)}=C(h,r) \frac{d \ln f}{d x},
\end{equation}
where $C(h,r)$ is a constant independent of the function $f$ but dependent of the kernel width $h$ and particle size $r$.

\subsection{Particle volume definitions}

Using the above approximations of convolution with the Dirac delta function, we may propose several distinct definitions of particle volumes. Namely, the choice $f=\rho$ gives
\begin{equation*}
  \rho(x_\alpha)=\rho_\alpha\approx \sum_\beta V_\beta \rho(x_\beta) W_{\alpha\beta}=\sum_\beta m_\beta W_{\alpha\beta},
\end{equation*}
and hence entails the above mass density update expression Eq. \eqref{eq.rhoa} or, generally, Eq. \eqref{eq.DefOfDensVar}. From this expression, one may define a particle volume (based on mass density) as
\begin{equation}\label{eq:Vm}
  (V^m_\alpha)^{-1}=\frac{\rho_\alpha}{m_\alpha} \approx \sum_\beta \frac{m_\beta}{m_\alpha} W_{\alpha\beta}.
\end{equation}

However, the introduction of another volume density variable, as is the case with entropy density, gives another possibility
\begin{equation}\label{eq:Vs}
  (V^s_\alpha)^{-1}=\frac{s_\alpha}{S_\alpha} \approx \sum_\beta \frac{S_\beta}{S_\alpha} W_{\alpha\beta},
\end{equation}
where we chose $f=s$ and where $S_{\sigma}$ is the entropy of particle $\alpha$.

At the same time, the choice $f:x\mapsto 1/V_{\alpha}$, $x\in V_{\alpha}$ gives directly an estimate of the particle volume as
\begin{equation} \label{eq:Vv}
  (V^{d}_\alpha)^{-1}\approx \sum_\beta W_{\alpha\beta}.
\end{equation}
Finally, the choice $f=\chi_{\Omega}$ being the characteristic function of a given domain $\Omega$ yields an implicit relation for particle volumes
\begin{equation*}
  1\approx \sum_\beta \tilde{V}^I_\beta W_{\alpha\beta},
\end{equation*}
where $\tilde{V}^{I}_{\alpha}$ is the volume of the part of the particle that is within the domain $\Omega$. This expression can be rearranged into
\begin{equation}
  \label{eq:VI}
  \tilde{V}_{\alpha}^{I} = \sum_{\beta} W_{\alpha\beta}^{-1},
\end{equation}
where $W_{\alpha\beta}^{-1}$ denotes the inverse to the matrix $W_{\alpha\beta}$.

Let us make few observations regarding the errors and convergence. All four definitions of volumes are based on Eulerian quantities whose evolution we aim to calculate. Hence, the error upper bound is only a rough guidance to the suitable volume definition selection as it is dependent on the systems evolution. Nevertheless, we expect from Eq. \eqref{eq:Error} that the mass-based volume should give a good particle volume representations in systems, where the spatial gradient of density is small and entropy-based volume should similarly yield a well-represented system with shallow entropy gradients. The direct definition of particle volume, $V^{d}_{\alpha}$, has errors proportional to the spatial variation of particle volumes and hence is dependent on the heterogeneity of discretization itself. Finally, the implicit definition of volume, $V^{I}_{\alpha}$, can be expected to have the smallest discretization error in the bulk, however, is expected to introduce errors in the boundary particles where there is a jump in the discretized function -- the characteristic function of the domain $\Omega$. In addition, the determination of implicit volume $V^{I}_{\alpha}$ comes at a greater computational cost as it involves inverting the large matrix $W_{\alpha\beta}$.

The precision of the implicit particle volumes $V^I_\alpha$ for the bulk integration can be perceived from the opposite perspective: we are looking for particle volumes such that the integral is exactly represented by the sum. Namely, the particle volumes $V^I_\alpha$ can be equivalently defined as weights of a Gaussian quadrature rule
\begin{equation*}
    \int f \mathrm{d} \xx \approx Q[f] = \sum_\alpha V^I_\alpha f(\xx_\alpha),
\end{equation*}
which is exact for all functions in the form
\begin{equation} \label{eq.fFormGaussQuad}
    f(\xx) = \sum_\beta f_\beta W(\xx - \xx_\beta)
\end{equation}
(where $f_\beta \in \mathbb{R}$ are arbitrary coefficients). Indeed, from the definition \eqref{eq:VI}:
\begin{equation*}
    Q[f] = \sum_\alpha V^I_\alpha f(\xx_\alpha) = \sum_\alpha \sum_\beta V^I_\alpha f_\beta W_{\alpha \beta} = \sum_\beta f_\beta = \sum_\beta f_\beta \int W(\xx - \xx_\beta) \mathrm{d} \xx = \int f(\xx) \mathrm{d} \xx.
\end{equation*}
To obtain the reverse implication, substitute $f(\xx) = W(\xx - \xx_\alpha)$ into $\int f \mathrm{d} \xx = Q[f]$. Note that again the issue with boundary elements appear, here in the requirement of the function form \eqref{eq.fFormGaussQuad}, which effectively requires $\Omega$ to be exactly discretized into SPH particles.

Let us, moreover, comment on the convergence of all the four above definitions of volume. The discretization of the convolution integral and Dirac delta function are improving, and hence the error being reduced, if the discretization step is decreased. At the same time, it is known that there is a tradeoff between the particle size and the SPH kernel width, reaching an optimum and preventing the truncation error to be arbitrarily small, \cite{quinlan}.

Yet another possibility is to define both mass and entropy densities as the SPH averages,
\begin{subequations}\label{eq.mixed}
	\begin{align}
		\rho_\alpha =& \sum_\beta m_\beta W_{\alpha\beta}\\
		s_\alpha =& \sum_\beta S_\beta W_{\alpha\beta}.
	\end{align}
\end{subequations}
This choice, called the mixed-volume approach, however, does not specify how a single particle volume is defined and instead needs two particle volumes at once. Despite this loss of elegance, it shows good results in the presence of discontinuities, see Section \ref{sec.jump}.

\subsection{Particle volumes and conservation laws}

We shall now turn to the discussion of the suitability of the particle volume definitions from a different perspective -- the validity of discretized conservation laws.

Using the mass-based volume $V^m_\alpha$, Eq.  \eqref{eq:Vm}, has the advantage of conserving the total mass, $M = \sum_\alpha V^m_\alpha \rho_\alpha = \sum_\alpha m_\alpha$, as particle masses $m_\alpha$ are constant.  Similarly, the total entropy, $S = \sum_\alpha V^m_\alpha s_\alpha = \sum_\alpha S_\alpha$, is conserved by the reversible part of the evolution (the Hamiltonian part, which does not affect $S_\alpha$).

The expression for total energy, Eq. \eqref{eq.SPH.ES}, however, differs according to the choice of the particle volume definition. In the mass-based volume case, it reads
 \begin{equation*}
     E^{SPH} = \sum_\alpha \frac{\MM^2_\alpha}{2 m_\alpha} +\sum_\alpha \frac{m_\alpha}{\rho_\alpha} \epsilon(\rho_\alpha, s_\alpha).
 \end{equation*}
If we consider the entropy-based particle volumes, the energy discretization takes on a slightly different form:
 \begin{equation*}
     E^{SPH} = \sum_\alpha \frac{\MM^2_\alpha}{2 m_\alpha} +\sum_\alpha \frac{S_\alpha}{s_\alpha} \epsilon(\rho_\alpha, s_\alpha).
   \end{equation*}

In the rest of this paper, we shall explore these four choices of particle volume definitions both in the reversible and irreversible processes, and discuss the conservation of energy. 


\subsection{Particle volumes and quadratures}
\begin{figure}[hbt!]
    \centering
    \includegraphics[width=0.6\linewidth]{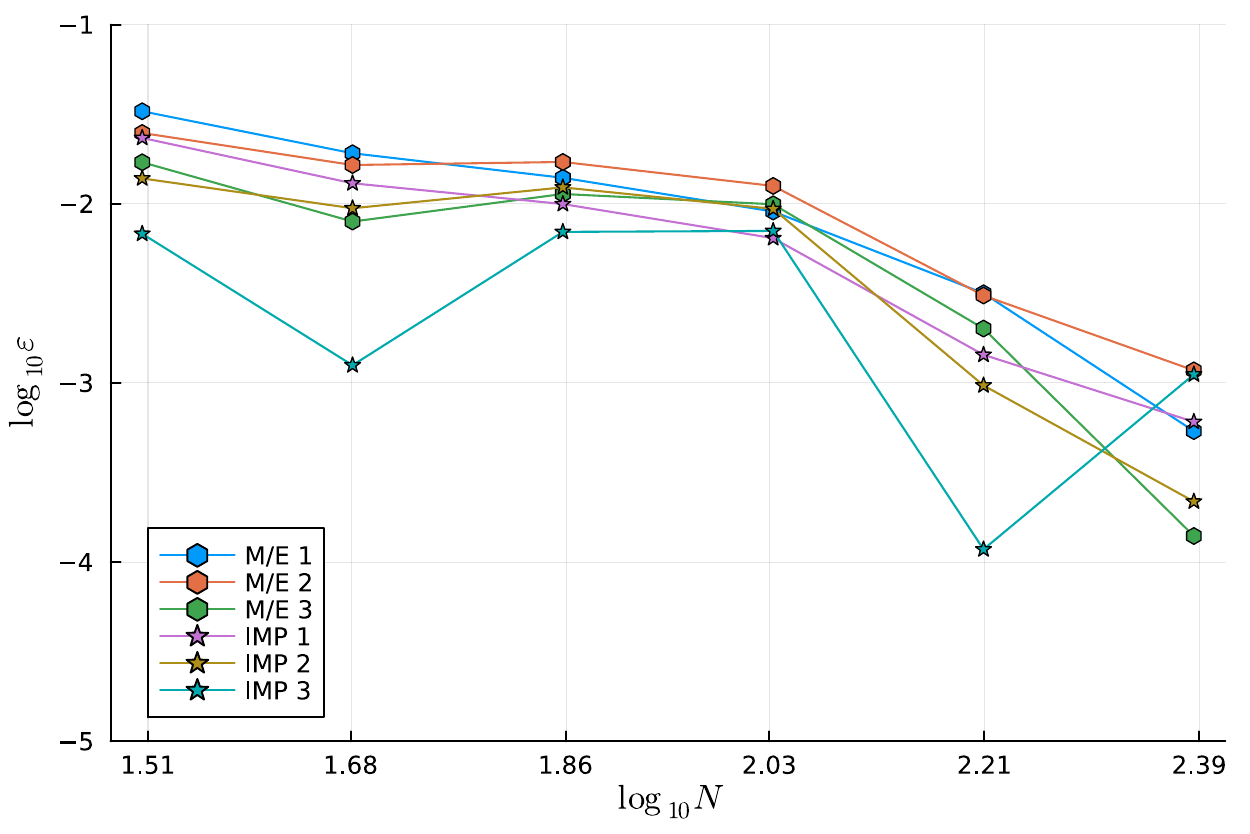}
    \caption{The relative accuracy of numerical quadrature $\epsilon$ for explicit integration using mass-volume or entropic-volume (M/E) and implicit-volume (IMP) for various values of $N$ and functions $f_1, f_2, f_3$ from \eqref{eq:testfuns}. (Note that in this example, mass-volume and entropic-volume yield identical results.) Dummy particles were used to prevent large error near boundary.}
    \label{fig.integration}
\end{figure}
As a purely mathematical comparison, let us investigate the accuracy of numerical integration
\begin{equation}
    \int f_1(\xx) \, \mathrm{d} \xx \approx \sum_a V_a f(\xx_a)
\end{equation}
for the different definitions of particle volume $V_a$. To this end, we consider functions
\begin{equation}
    \begin{split}
        f_1 &= e^{-x-y} \\
        f_2 &= 1 + \cos 8x + \sin 8x \\
        f_3 &= 1 - xy + 4x^2y^2
    \end{split}
    \label{eq:testfuns}
\end{equation}
in a square $[0,1] \times [0,1]$. The exact integral can be easily evaluated. We approximate this value by $N$ particles arranged in a hexagonal pattern. Let us define the spatial resolution of the problem as
\begin{equation}
    \mathrm{d}r = \frac{1}{\sqrt{N}}.
\end{equation}
The particle masses $m_\alpha$ and entropies $S_\alpha$ were all set equal to 1 and Wendland's quintic kernel with support radius $h = 3 \mathrm{d}r$ was used. The result is shown in Figure \ref{fig.integration}. We found that the implicit volume is more precise in most cases (at the expense of a longer computation time). Therefore, we suggest that the implicit-volume quadrature should be used in a situation where an integral has to be computed from data in a post-process.


\subsection{Particle volumes in the presence of discontinuities}\label{sec.jump}
Let us now illustrate the difference between the volumes in the presence of a discontinuity in the mass density while keeping the entropy density constant in space. We consider a one-dimensional system of particles and make the particles twice denser in the right part of the domain (interval $x\in (0,1)$ split into two halves). All particles have the same mass, the interparticle spacing is $1.0\cdot 10^{-04}$ in the left part of the interval while $0.5\cdot 10^{-04}$ in the right part. Moreover, we set the SPH kernel size as $h = 0.001$ (using the Wendland 1D kernels). Therefore, the density of the particles in the left part should be approximately 1.0 while 2.0 in the right part, experiencing a jump in the middle. The entropies of the particles in the left part are set equal to the respective interparticle spacings so that the entropy is approximately constant over the whole domain. 

\begin{figure}[ht!]
\centering
\includegraphics[scale=0.5]{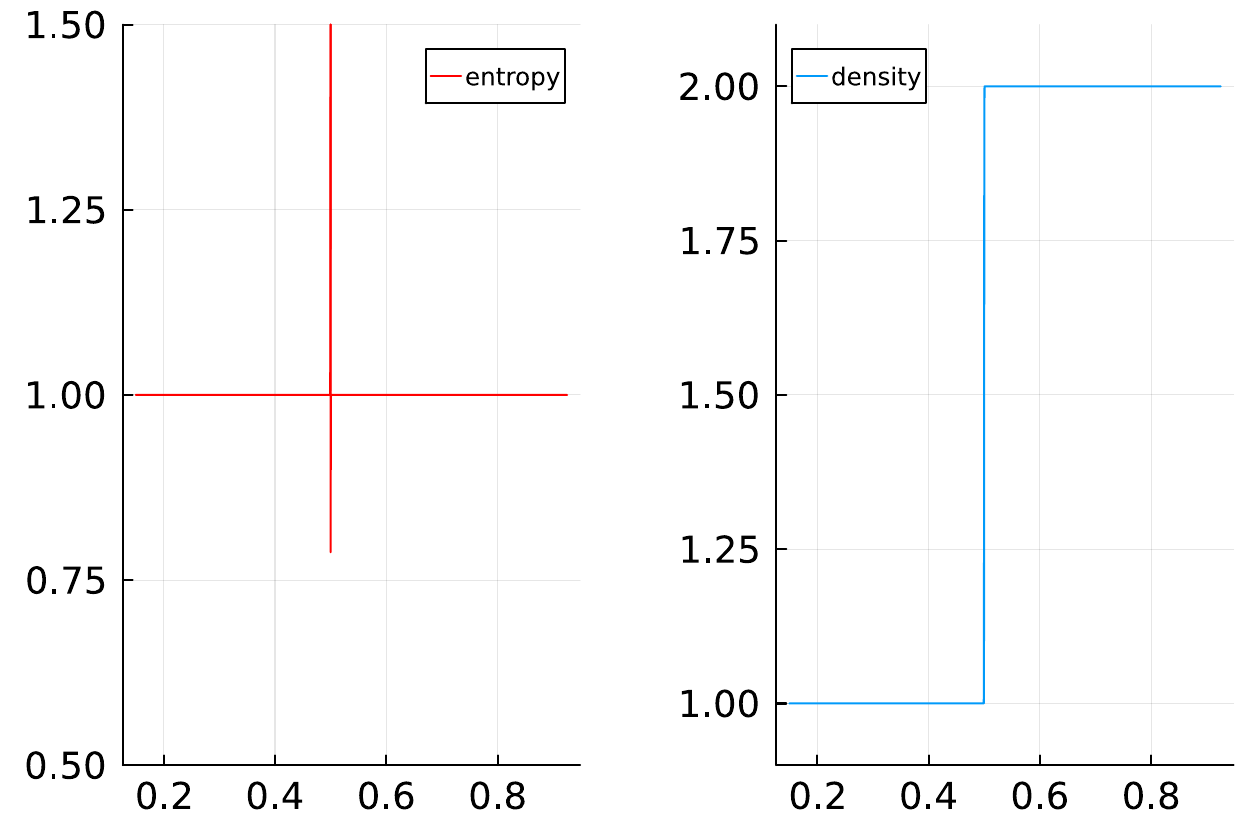}
\caption{\label{fig.de.mass}Entropy density and mass density in the presence of the mass density jump evaluated using the mass volume.}
\end{figure}

\begin{figure}[ht!]
\centering
\includegraphics[scale=0.5]{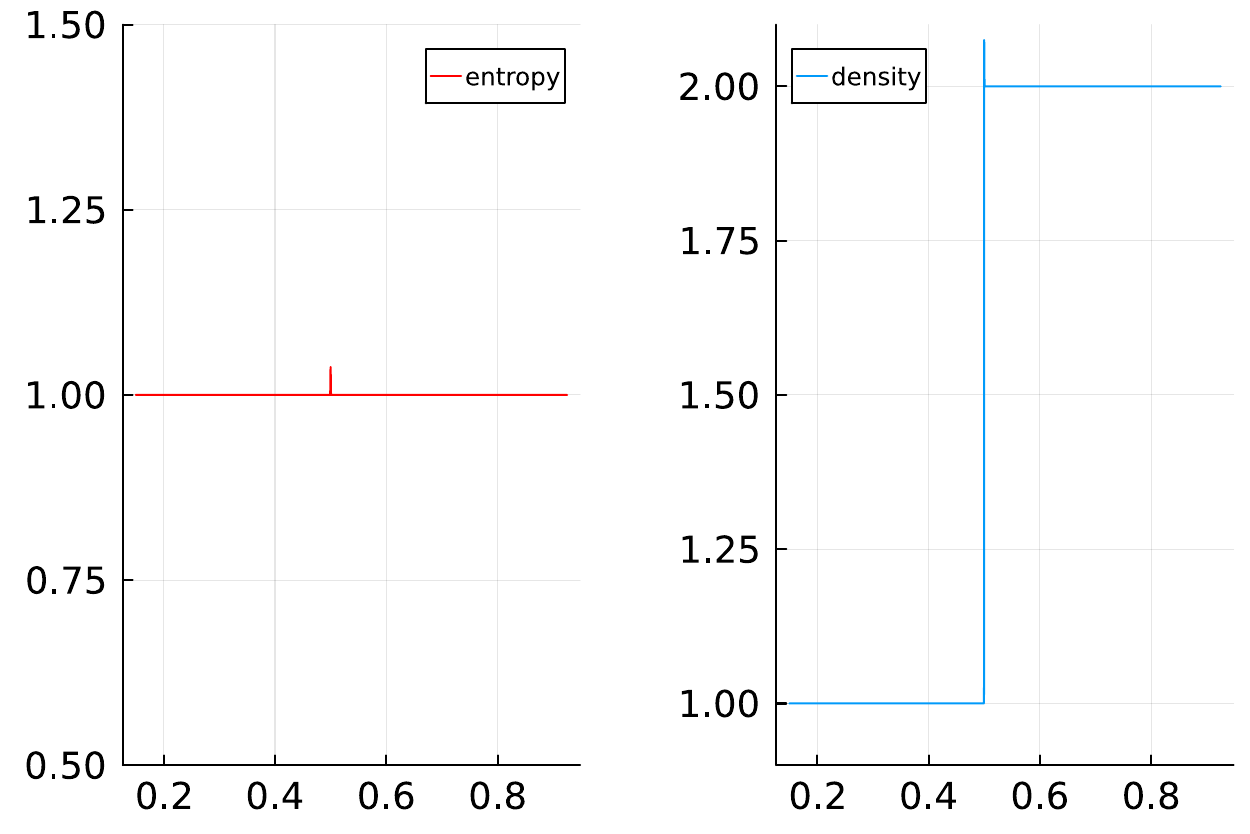}
\caption{\label{fig.de.entropic}Entropy density and mass density in the presence of the mass density jump evaluated using the entropic volume.}
\end{figure}

\begin{figure}[ht!]
\centering
\includegraphics[scale=0.5]{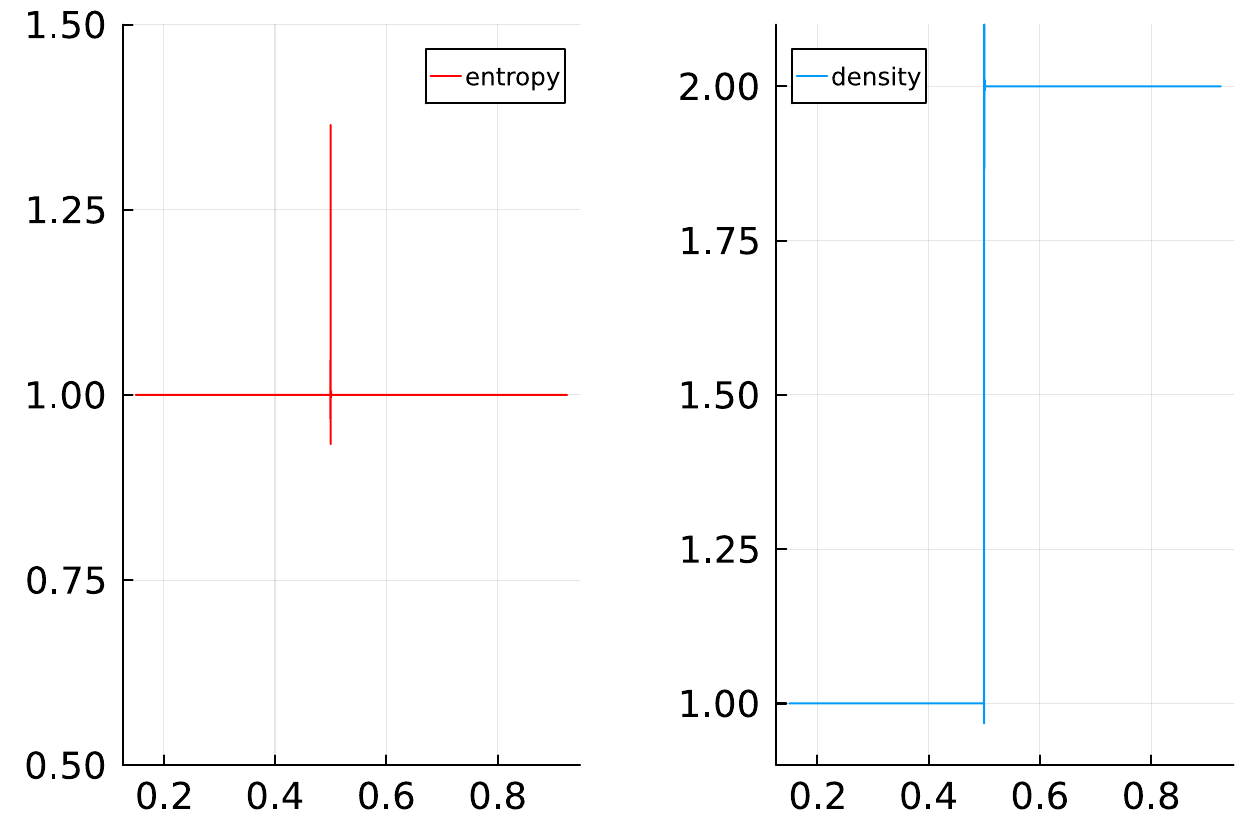}
\caption{\label{fig.de.implicit}Entropy density and mass density in the presence of the density jump evaluated using the implicit volume.}
\end{figure}

\begin{figure}[ht!]
\centering
\includegraphics[scale=0.5]{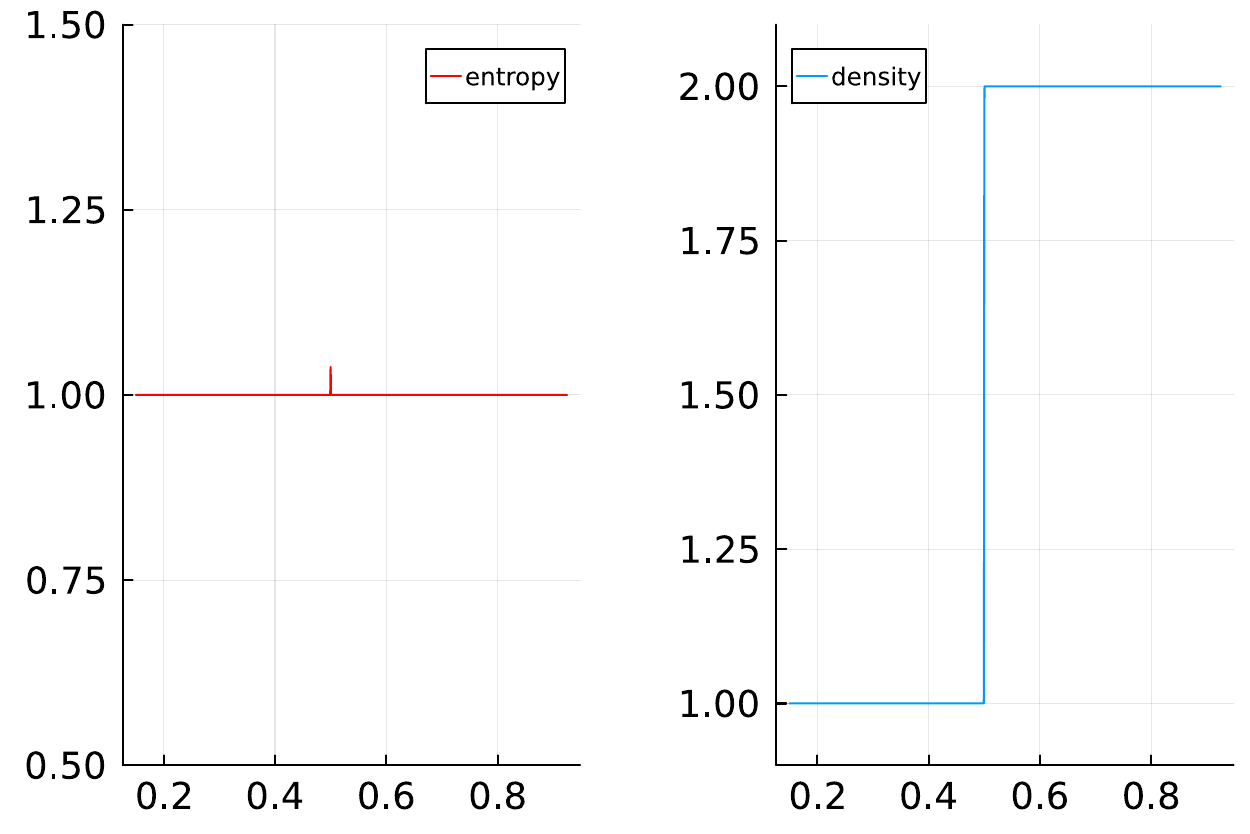}
\caption{\label{fig.de.mixed}Entropy density and mass density in the presence of the density jump evaluated using the mixed volume. This method shows the best behavior, capturing the mass density jump well while having a relatively small oscillation in the entropy density.}
\end{figure}

\begin{figure}[ht!]
\centering
\includegraphics[scale=0.5]{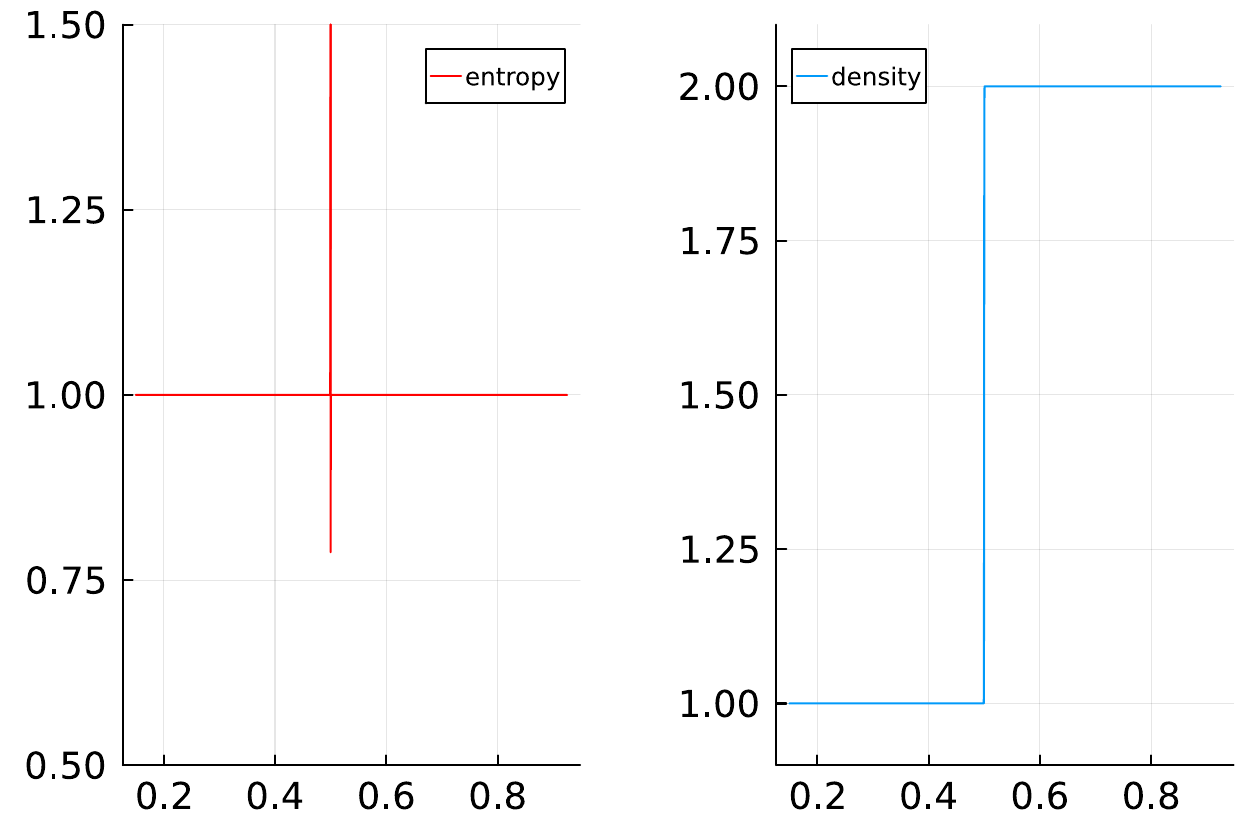}
\caption{\label{fig.de.direct}Entropy density and mass density in the presence of the density jump evaluated using the direct volume.}
\end{figure}

Figure \ref{fig.de.mass} shows the entropy density and the mass density evaluated using the mass-volume method. While the mass density shows a smooth monotone profile, the entropy density oscillates considerably. On the other hand, when we use the entropic-volume discretization, the entropy gets a bump in the middle (caused by oversampling the particles to the right of the jump), as well as the mass density; see Figure \ref{fig.de.entropic}. The implicit volume shows oscillations in both the mass density and the entropy density, see Figure \ref{fig.de.implicit}. The direct-volume approach is shown in Figure \ref{fig.de.direct}. Finally, Figure \ref{fig.de.mixed} shows the behavior of the mixed volume, which gives the best results. 

In summary, it may be advantageous to replace the standard mass volume with the entropic volume when the entropy has a smoother behavior than the mass density. In the following Section, we show the reversible part of the SPH evolution equations for each of those SPH variants.

\section{Hamiltonian part of the SPH evolution with entropy}\label{sec.ham.ent}
Since the reversible part of the SPH evolution can be seen as a discretization of the Lagrangian continuum mechanics, it is generated by a Poisson bracket. However, the five forms of SPH with entropy discussed in this paper have different Poisson brackets and different reversible parts of the evolution equations. As in the case of standard barotropic SPH, Poisson brackets are derived by projection from the Lagrangian Poisson bracket \eqref{eq.PB.Lag}. The projection consists of the mapping from the Lagrangian state variables $\xx(\XX)$ and $\MM(\XX)$ to the SPH position $\xx_\alpha$ and $\MM_\alpha$ by Equations \eqref{eq.projection}, and of the respective definitions of the mass density and entropy density based on the choice of the definition of the discrete volume.

\subsection{Mass-based volume approach}
The Poisson bracket governing kinematics of $\xx_\alpha$, $\MM_\alpha$ (given by Equations \eqref{eq.projection}), mass density $\rho_\alpha$, and the volumetric entropy density $s_\alpha$ (given by Equations \eqref{eq:LagEulStateVars}), that is
\begin{subequations}
	\begin{align}
		\rho_\alpha =& \sum_\beta m_\beta W_{\alpha\beta}\\
		s_\alpha =& \frac{S_\alpha \rho_\alpha}{m_\alpha},
	\end{align}
\end{subequations}
is obtained by plugging these projections into the Lagrangian Poisson bracket \eqref{eq.PB.Lag}. Note that $S_\alpha$ represents the entropy of the SPH particle $\alpha$. Appendix \ref{sec.PB.m} contains details of the calculation. The resulting Hamiltonian evolution equations are
\begin{subequations}  \label{eq.extSPH}
    \begin{align}
        \dot{\xx}_\alpha &= E_{\MM_\alpha}\\
        \dot{\MM}_\alpha &= -E_{\xx_\alpha} - \sum_\beta m_\alpha m_\beta \left(\frac{E_{\rho_\alpha}}{m_\alpha} +\frac{E_{\rho_\beta}}{ m_\beta}+S_\alpha \frac{1}{m_\alpha^2} E_{s_\alpha}+S_\beta\frac{1}{m_\beta^2} E_{s_\beta}\right)W'_{\alpha\beta}\ee_{\alpha\beta}\\
        \dot{\rho}_\alpha &= \sum_\beta m_\beta W'_{\alpha\beta}\ee_{\alpha\beta}\cdot(E_{\MM_{\alpha}}-E_{\MM_\beta})\\
        \dot{s}_\alpha &= \frac{S_\alpha}{m_\alpha}\sum_\beta m_\beta W'_{\alpha\beta}\ee_{\alpha\beta}\cdot(E_{\MM_{\alpha}}-E_{\MM_\beta})
    \end{align}
\end{subequations} 
These equations are an extension of the standard SPH equations \eqref{eq.SPH.standard} with an explicit evolution of $\rho_\alpha$ of $s_\alpha$.
Note that $\dot{\rho}_\alpha$ in (\ref{eq.extSPH}c) is directly compatible with the ``update'' relation for $\rho_\alpha$ (Equation \eqref{eq.rhoa}) as one can check by differentiating and using (\ref{eq.extSPH}a), and if the update relation is used instead of the evolution equation for $\rho_\alpha$, the system of evolution equations becomes symplectic. 

To close Equations \eqref{eq.extSPH}, we need to supply an energy functional as a function of the state variables $\xx_{\alpha}$, $\MM_{\alpha}$, $\rho_{\alpha}$, and $s_{\alpha}$, which in particular can also depend on the entropy. 
Equations \eqref{eq.extSPH} then turn to 
\begin{subequations}\label{eq.SPH.m.evo}
    \begin{align}
      \dot{x}^i_\alpha &= \frac{M^i_\alpha}{m_\alpha}\\
      \dot{\rho}_\alpha &= \sum_\beta m_\beta W'_{\alpha\beta}\ee_{\alpha\beta}\cdot\left(\frac{\MM_\alpha}{m_\alpha} - \frac{\MM_\beta}{m_\beta}\right)\\
      \dot{M}_{\alpha i} &=  -\sum_\beta m_\alpha m_\beta\left(\frac{p_\alpha}{\rho^2_\alpha} + \frac{p_\beta}{\rho^2_\beta} \right) W'_{\alpha\beta}e_{\alpha \beta i}\\
        \dot{s}_\alpha &= \frac{S_\alpha}{m_\alpha}\sum_\beta m_\beta W'_{\alpha\beta}\ee_{\alpha\beta}\cdot\left(\frac{\MM_\alpha}{m_\alpha} - \frac{\MM_\beta}{m_\beta}\right),
    \end{align}
where pressure and temperature are defined as
  \end{subequations}
\begin{equation}\label{eq.p.S}
    p_\alpha = -\epsilon + \rho_\alpha \frac{\partial \epsilon}{\partial \rho_\alpha}+ s_\alpha \frac{\partial \epsilon}{\partial s_\alpha}
    \quad\mbox{and}\quad
    T_\alpha = \frac{\partial \epsilon}{\partial s_\alpha},
\end{equation}
and where $\vv_{\alpha} = \frac{\MM_\alpha}{m_\alpha}$ is the particle velocity.

Note that $E_{\xx_\alpha}=0$ for each $\alpha$ for energy \eqref{eq.SPH.ES} with $V_{\alpha}=V_{\alpha}^{m}$ and that the evolution equations for position and momentum are identical while the evolution equations for particle density and entropy are equivalent to their updated version. As noted above, the total mass and entropy are conserved, since the Hamiltonian evolution does not change the particle entropies, $S_\alpha$.

\subsection{Entropic-based volume approach}\label{sec.entropic}
Also with volume defined via the ratio of the entropy density and entropy of each particle $V_{\alpha}^{s}$, the SPH evolution equations can be derived by reduction from the Poisson bracket of Lagrangian continuum mechanics \eqref{eq.PB.Lag}. The resulting Hamiltonian evolution equations are analogical to Equations \eqref{eq.extSPH} with the roles of $\rho_\alpha$ and $m_\alpha$ swapped with $s_\alpha$ and $S_\alpha$, that is
\begin{subequations}\label{eq.s.rho.entropic}
	\begin{align}
		s_\alpha =& \sum_\beta S_\beta W_{\alpha\beta}\\
		\rho_\alpha =& \frac{m_\alpha s_\alpha}{S_\alpha},
	\end{align}
\end{subequations}
and
\begin{subequations}  \label{eq.extSPH.s}
    \begin{align}
        \dot{\xx}_\alpha &= E_{\MM_\alpha}\\
        \dot{\MM}_\alpha &= -E_{\xx_\alpha} - \sum_\beta S_\alpha S_\beta \left(\frac{m_\alpha}{S_\alpha^2} E_{\rho_\alpha} + \frac{m_\beta}{S_\beta} E_{\rho_\beta}+\frac{1}{S_\alpha} E_{s_\alpha}+\frac{1}{S_\beta} E_{s_\beta}\right)W'_{\alpha\beta}\ee_{\alpha\beta}\\
        \dot{s}_\alpha &= \sum_\beta S_\beta W'_{\alpha\beta}\ee_{\alpha\beta}\cdot (E_{\MM_{\alpha}}-E_{\MM_\beta})\\
        \dot{\rho}_\alpha &= \frac{m_\alpha}{S_\alpha}\sum_\beta S_\beta W'_{\alpha\beta}\ee_{\alpha\beta}\cdot (E_{\MM_{\alpha}}-E_{\MM_\beta}).
    \end{align}
\end{subequations} 

In addition, these equations can be made symplectic by keeping the evolutions of $\xx_\alpha$ and $\MM_\alpha$ while evaluating $s_\alpha$ and $\rho_\alpha$. 
Taking energy \eqref{eq.SPH.ES} with $V_{\alpha}=V_{\alpha}^{s}$, the evolution equations simplify to 
\begin{subequations}\label{eq.FM.s}
    \begin{align}
      \dot{x}^i_\alpha &= \frac{M^i_\alpha}{m_\alpha}\\
        \dot{M}_{\alpha i} &= -\sum_\beta S_\alpha S_\beta \left(\frac{p_\alpha}{s_\alpha^2}+\frac{p_\beta}{s_\beta^2}\right) W'_{\alpha\beta} e_{\alpha \beta i}\\
        \dot{s}_\alpha &= \sum_\beta S_\beta W'_{\alpha\beta}\ee_{\alpha\beta}\cdot(\vv_\alpha-\vv_\beta)\\
        \dot{\rho}_\alpha &= \frac{m_\alpha}{S_\alpha}\sum_\beta S_\beta W'_{\alpha\beta}\ee_{\alpha\beta}\cdot(\vv_\alpha-\vv_\beta),
    \end{align}
\end{subequations}
which can again be made symplectic when both the entropy density and mass density are calculated directly via relations \eqref{eq.s.rho.entropic}.

\subsection{Direct volume approach}

The definition of the direct volume is
\begin{equation} \label{eq:Vd}
  V^{d}_\alpha=\left(\sum_\beta W_{\alpha\beta} \right)^{-1}.
\end{equation}
The density state variables simply follow from the evolution of direct volume $V^d_\alpha$ expression and their Lagrange initial values:
\begin{equation}
\rho_\alpha=m_\alpha/V^d_\alpha, \quad s_\alpha = S_\alpha/V^d_\alpha,
\end{equation}
where $\rho_{\alpha}$ and $s_{\alpha}$ are independent state variables (once dissipative evolution is considered).

The corresponding Poisson bracket is the same as in the mass-based volume approach, see Appendix \ref{sec.PB.m}, but where $m_\beta$ is replaced by $m_\alpha$. The resulting Hamiltonian evolution equations are
\begin{subequations}  \label{eq.extSPHd}
    \begin{align}
        \dot{\xx}_\alpha &= E_{\MM_\alpha}\\
        \dot{\MM}_\alpha &= -E_{\xx_\alpha} - \sum_\beta \left(m_\alpha E_{\rho_\alpha} +m_\beta E_{\rho_\beta}+S_\alpha E_{s_\alpha}+S_\beta E_{s_\beta}\right)W'_{\alpha\beta}\ee_{\alpha\beta}\\
        \dot{\rho}_\alpha &= m_\alpha \sum_\beta  W'_{\alpha\beta}\ee_{\alpha\beta}\cdot(E_{\MM_{\alpha}}-E_{\MM_\beta})\\
        \dot{s}_\alpha &= S_\alpha \sum_\beta W'_{\alpha\beta}\ee_{\alpha\beta}\cdot(E_{\MM_{\alpha}}-E_{\MM_\beta})
    \end{align}
\end{subequations}

Using the considered form of energy \eqref{eq.SPH.ES}, equations \eqref{eq.extSPHd} then turn to 
\begin{subequations}\label{eq.SPH.d.evo}
    \begin{align}
      \dot{x}^i_\alpha &= \frac{M^i_\alpha}{m_\alpha}\\
      \dot{\rho}_\alpha &= m_\alpha \sum_\beta W'_{\alpha\beta}\ee_{\alpha\beta}\cdot\left(\frac{\MM_\alpha}{m_\alpha} - \frac{\MM_\beta}{m_\beta}\right)\\
      \dot{M}_{\alpha i} &=  -\sum_\beta \left[\left(\frac{m_\alpha}{\rho_\alpha}\right)^2 p_\alpha + \left(\frac{m_\beta}{\rho_\beta}\right)^2 p_\beta\right] W'_{\alpha\beta}e_{\alpha \beta i}\\
        \dot{s}_\alpha &= S_\alpha\sum_\beta W'_{\alpha\beta}\ee_{\alpha\beta}\cdot\left(\frac{\MM_\alpha}{m_\alpha} - \frac{\MM_\beta}{m_\beta}\right),
    \end{align}
where pressure and temperature are defined as in Equation \eqref{eq.p.S}.
\end{subequations}

\subsection{Implicit-based volume approach}\label{sec.implicit}



The definition of implicit-based volume relates mass and entropy density as
\begin{equation}
  \label{eq:VaI}
  V_\alpha^I = \frac{m_\alpha}{\rho_\alpha} = \frac{S_\alpha}{s_\alpha} = \sum_{\beta} W_{\alpha\beta}^{-1}
\end{equation}
but note that both $\rho_{\alpha}$ and $s_{\alpha}$ are independent state variables as we shall consider dissipative evolution as well, where we let $S_{\alpha}$ to evolve to reflect the dissipation. Nevertheless, the evolution of eulerian entropic density $s_{\alpha}$ is such that the above relation Eq. \eqref{eq:VaI} for implicit particle volume holds. 

The Poisson bracket governing the reversible evolution of $\xx_\alpha$, $\MM_\alpha$, $\rho_\alpha$, and $s_\alpha$ can be obtained by plugging functionals dependent on these state variables into the Poisson bracket of Lagrangian continuum mechanics \eqref{eq.PB.Lag}, see Appendix \ref{sec.PB.I}. We again assume that the Lagrangian particle mass and entropy are independent of the particle positions (although $S_{\alpha}$ can change over time due to irreversible effects).  The reversible part of the evolution equations then reads
\begin{subequations}\label{eq.imp.rev}
\begin{align}
\dot{\xx}_\alpha &= E_{\MM_\alpha}\\
  \dot{\rho}_\alpha &= \frac{\rho_\alpha^2}{m_\alpha} \sum_\beta \frac{m_\beta}{\rho_\beta} \sum_\delta W^{-1}_{\alpha\delta}W'_{\delta\beta}\ee_{\delta\beta}\cdot(E_{\MM_\delta} - E_{\MM_\beta})\\
  \dot{\MM}_\alpha &= -E_{\xx_\alpha} 
- \sum_\beta\left(\frac{m_\beta}{\rho_\beta}(E^{-1}_\rho)_\alpha+ \frac{m_\alpha}{\rho_\alpha}(E^{-1}_\rho)_\beta
 +\frac{S_\beta}{s_\beta}(E^{-1}_s)_\alpha+ \frac{S_\alpha}{s_\alpha}(E^{-1}_s)_\beta\right)W'_{\alpha\beta}\ee_{\alpha\beta}\\ 
\dot{s}_\alpha &= \frac{s_\alpha^2}{S_\alpha} \sum_\beta \frac{S_\beta}{s_\beta} \sum_\delta W^{-1}_{\alpha\delta}W'_{\delta\beta}\ee_{\delta\beta}\cdot(E_{\MM_\delta} - E_{\MM_\beta}),
\end{align}
\end{subequations}
where $(E^{-1}_\rho)_\alpha = \sum_\gamma E_{\rho_\gamma}\frac{\rho^2_\gamma}{m_\gamma} W^{-1}_{\gamma\alpha}$
and $(E^{-1}_s)_\alpha = \sum_\gamma E_{s_\gamma}\frac{s^2_\gamma}{S_\gamma} W^{-1}_{\gamma\alpha}$. Note that in the case where $m_{\alpha}$ and $S_{\alpha}$ are constants (the reversible case), the evolution equations for $\rho_{\alpha}$ and $s_{\alpha}$, Equations (\ref{eq.imp.rev}b,d), can be rewritten in terms $V_{\alpha}^{I}$ as expected.

Equations \eqref{eq.imp.rev} represent a Hamiltonian system of ODE's that is non-symplectic. Indeed, the total mass and total entropy are preserved regardless of the choice of energy, so the underlying Poisson bracket has Casimir functionals (mass and entropy), that are conserved regardless of the choice of energy. The system of equations can be made symplectic as before by explicitly evaluating the entropy and mass density by updating the rule \eqref{eq:LagEulStateVars} (while removing the equations $\dot{\rho}_{\alpha}$ and $\dot{s}_{\alpha}$ from \eqref{eq.imp.rev}). With energy $E = \sum_\alpha \left(\frac{\MM^2_\alpha}{2m_\alpha} + V^{I}_\alpha \epsilon(\rho_\alpha,s_\alpha)\right)$, symplectified Equations \eqref{eq.imp.rev} become
\begin{subequations}\label{eq.imp.rev.sym}
\begin{align}
\dot{\xx}_\alpha &= E_{\MM_\alpha}\\
\dot{\MM}_\alpha &= -\sum_\gamma\sum_\beta p_\gamma(V_\beta W^{-1}_{\gamma\alpha} + V_\alpha W^{-1}_{\gamma\beta}) W'_{\alpha\beta}\ee_{\alpha\beta}.
\end{align}
\end{subequations}

However, there is a complication in the numerical implementation of Equations \eqref{eq.imp.rev.sym} caused by the presence of the inverse matrix $W_{\alpha\beta}^{{-1}}$. While the matrix $W_{\alpha\beta}$ is typically sparse, its inverse is dense and therefore it is impractical to store it in memory. We can go around this obstacle by solving the linear system of equations $\sum_\beta  W_{\gamma\beta} y_\beta= p_\gamma$ 
at each time step by conjugate gradients, since the solution gives us the terms on the right-hand side of the momentum equations, $y_\alpha = p_\gamma W^{-1}_{\gamma\alpha}$. Although this approach slows down the simulation (we are using explicit time stepping to keep the symplecticity), the simulation is still viable on standard desktops (for $10^4$ particles in two dimensions). 

\subsection{Mixed-volume approach}
Finally, another possibility is to define the mass and entropy densities using Equations \eqref{eq.mixed},
which mixes the mass-based particle volume with the entropy-based volume. Despite leaving the elegance of having a single-particle volume, the mixed approach provides smoothing to both the fields, which might be advantageous. 

Poisson bracket \eqref{eq.PB.x}, derived in Appendix \ref{sec.PB.x}, leads to the following evolution equations for density and entropy density defined through \eqref{eq.mixed},
\begin{subequations}  \label{eq.extSPH-x}
    \begin{align}
        \dot{\xx}_\alpha &= E_{\MM_\alpha}\\
        \dot{\MM}_\alpha &= -E_{\xx_\alpha} - \sum_\beta (m_\alpha E_{\rho_\beta} + m_\beta E_{\rho_\alpha}+S_\alpha E_{s_\beta}+S_\beta E_{s_\alpha})W'_{\alpha\beta}\ee_{\alpha\beta}\\
        \dot{\rho}_\alpha &= \sum_\beta m_\beta W'_{\alpha\beta}\ee_{\alpha\beta}(E_{\MM_{\alpha}}-E_{\MM_\beta})\\
        \dot{s}_\alpha &= \sum_\beta S_\beta W'_{\alpha\beta}\ee_{\alpha\beta}(E_{\MM_{\alpha}}-E_{\MM_\beta}),
    \end{align}
\end{subequations} 
where energy is yet to be supplied. Using the energy \eqref{eq.SPH.ES} with the mass volume, evolution equations \eqref{eq.extSPH-x} become
\begin{subequations}\label{eq.SPH.evo.mixed}
    \begin{align}
      \dot{x}^i_\alpha &= \frac{M^i_\alpha}{m_\alpha}\\
      \dot{\rho}_\alpha &= \sum_\beta m_\beta W'_{\alpha\beta}\ee_{\alpha\beta}\left(\frac{\MM_\alpha}{m_\alpha} - \frac{\MM_\beta}{m_\beta}\right)\\
        \dot{M}_{\alpha i} &= -\sum_\beta\left[\frac{m_\alpha m_\beta}{\rho^2_\alpha}p_\alpha + \frac{m_\alpha m_\beta}{\rho^2_\beta}p_\beta
        +m_\alpha m_\beta\left(\left(\frac{S_\beta}{m_\beta}-\frac{s_\alpha}{\rho_\alpha}\right)\frac{T_\alpha}{\rho_\alpha}+\left(\frac{S_\alpha}{m_\alpha}-\frac{s_\beta}{\rho_\beta}\right)\frac{T_\beta}{\rho_\beta}\right)\right] W'_{\alpha\beta}\ee_{\alpha \beta}\\
      \dot{s}_\alpha &= \sum_\beta S_\beta W'_{\alpha\beta}\ee_{\alpha\beta}\left(\frac{\MM_\alpha}{m_\alpha} - \frac{\MM_\beta}{m_\beta}\right),
    \end{align}
where temperature is defined as $T_\alpha = \frac{\partial \epsilon}{s_\alpha}$. Note the extra terms on the right-hand side of the evolution equation for $\MM_\alpha$, which depend on the entropies. Although these terms are not present in the usual form of SPH equations, they are necessary to maintain the Hamiltonianity of the equations when the mixed-volume approach is taken, and they follow directly from the projection of Poisson brackets. 
  \end{subequations}

\subsection{Illustration - adiabatic expansion}   
In this example, we consider adiabatic expansion, where the entropy of individual particles $S_\alpha$ does not change. This example can be seen as a baseline numerical experiment, as it is an example of a reversible process. We shall use it for discussing the effects of the different choices of particle volumes introduced above.

\begin{figure}[ht!]
\centering
\begin{tabular}{cc}
\subf{\includegraphics[width=60mm]{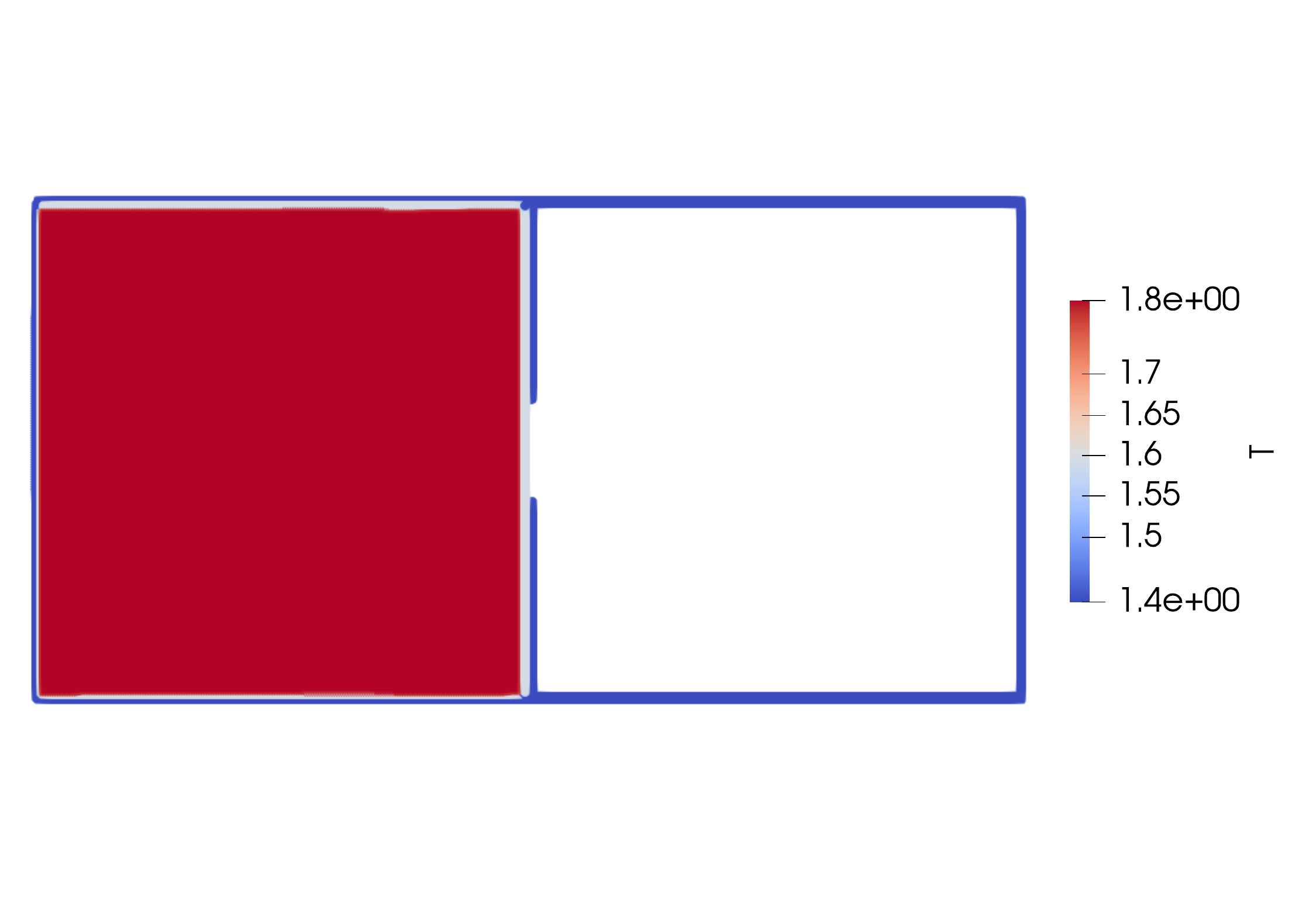}}
     {$t=0.0$}
&
\subf{\includegraphics[width=60mm]{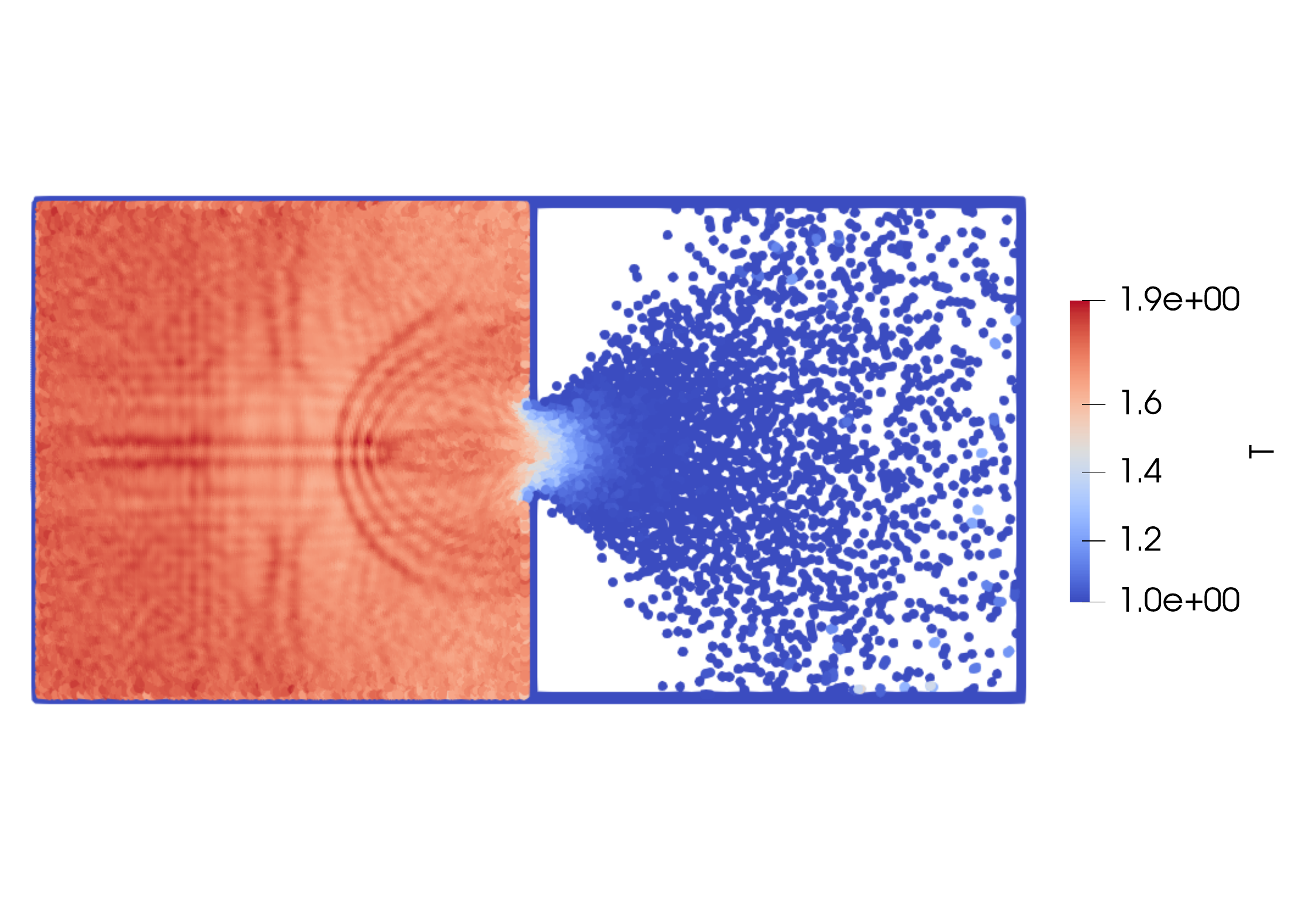}}
     {$t=0.6$}
\\
\subf{\includegraphics[width=60mm]{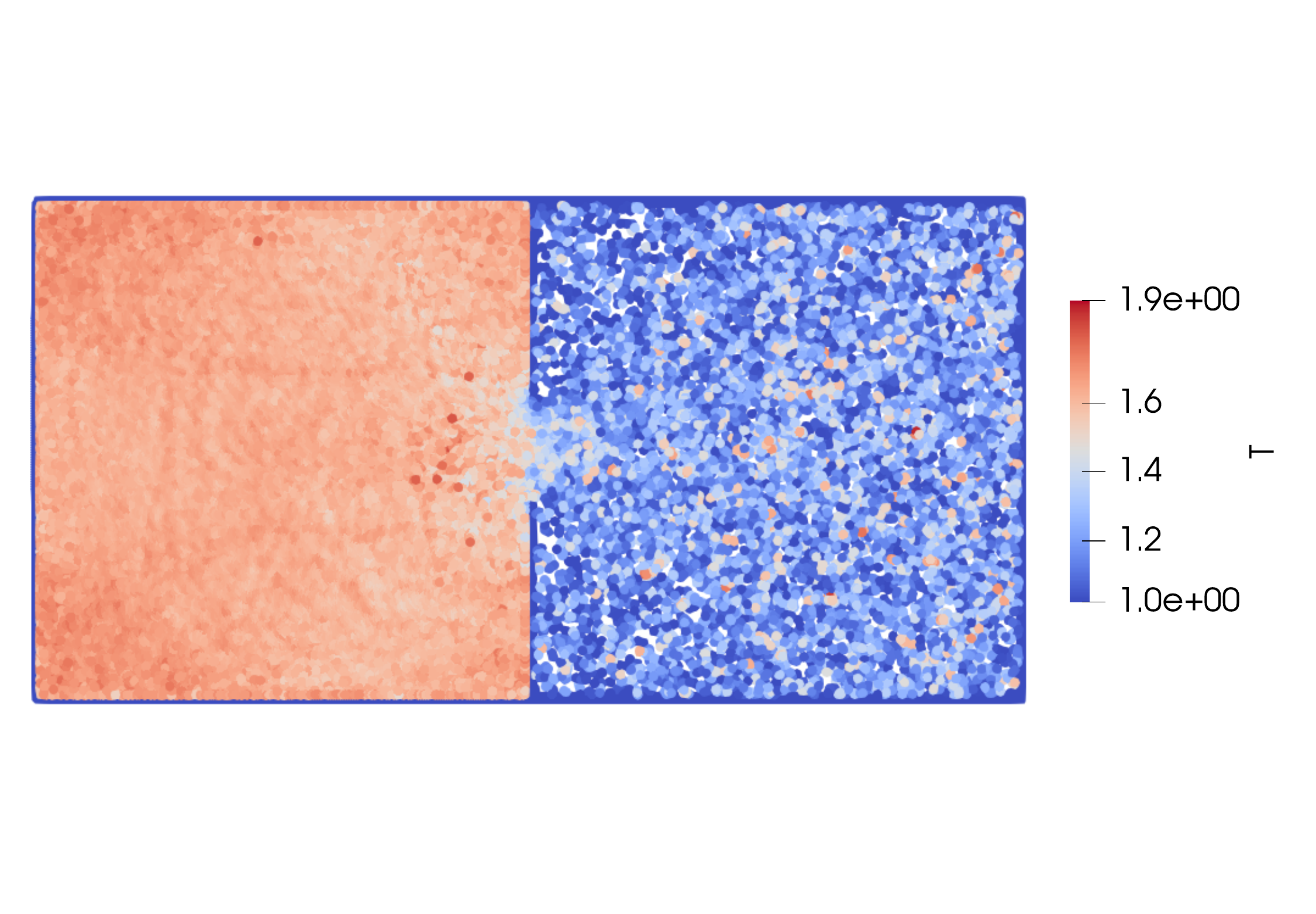}}
     {$t=2.5$}
&
\subf{\includegraphics[width=60mm]{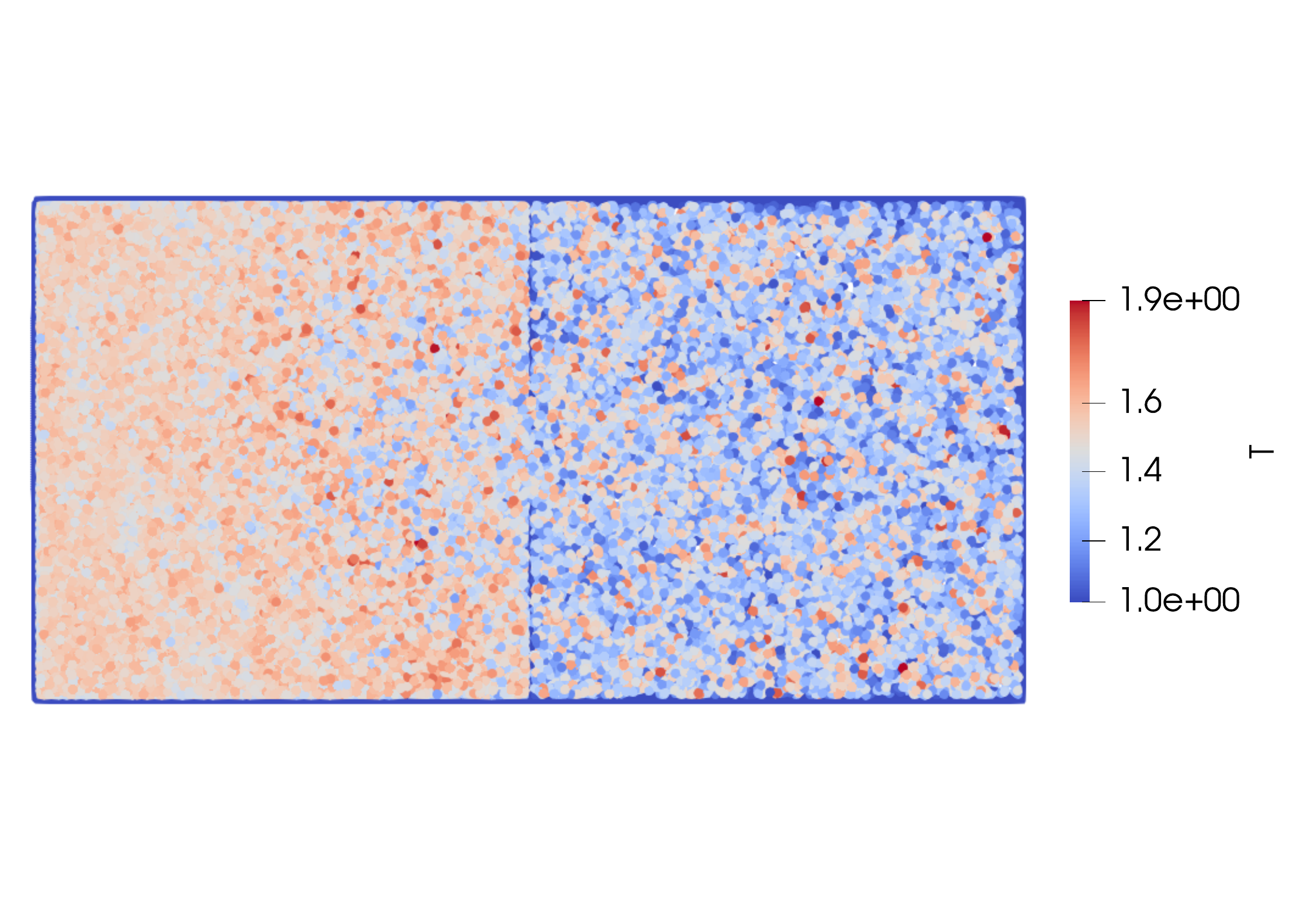}}
     {$t=15.0$}
\end{tabular}
\caption{\label{fig.adiabatic}Reversible simulation of adiabatic expansion of an ideal gas (with 43017 SPH particles, the Verlet integrator \cite{verlet}, final time 15, time step 1.3E-05, heat capacity $c_V=1.0J/kg K$, $\gamma=1.4$). The color indicates the temperature. The error in the total energy is $10^{-4}\%$ and decreases quadratically with the time step. The simulation is carried out with the mass-based particle volume. Simulations with the entropic, direct, and mixed particle volumes give very similar results. Simulation with the implicit volume leads to negative particle volumes due to boundary effects, so it is not shown.}
\end{figure}

We consider a double-compartment vessel with only the left half initially filled with an ideal gas; see Appendix \ref{sec.thermo} for the thermodynamics of the ideal gas. The whole vessel is eventually occupied and the temperature of the gas is reduced in accordance with the laws governing adiabatic processes Figure \ref{fig.adiabatic}.


\begin{figure}[ht!]
\centering
\begin{tabular}{cc}
\subf{\includegraphics[width=60mm]{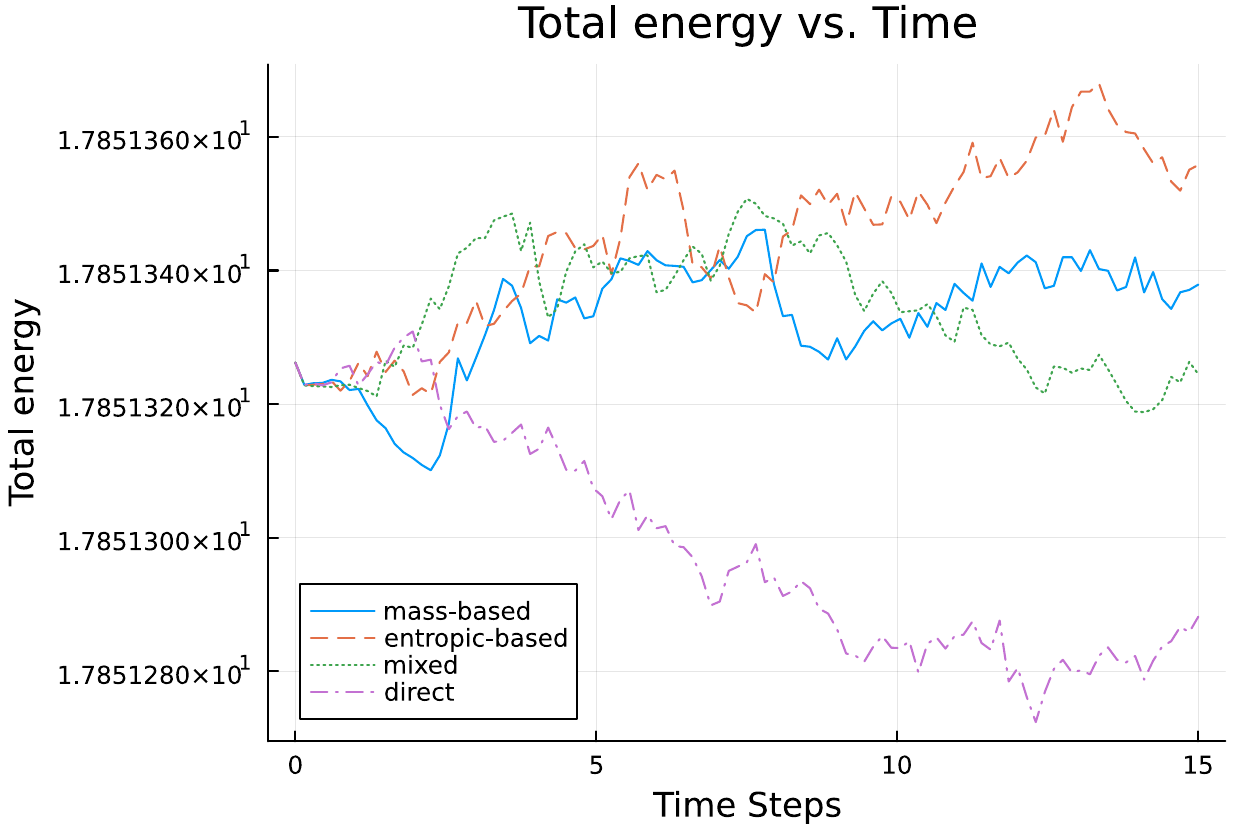}}
     {total energy in time}
&
\subf{\includegraphics[width=60mm]{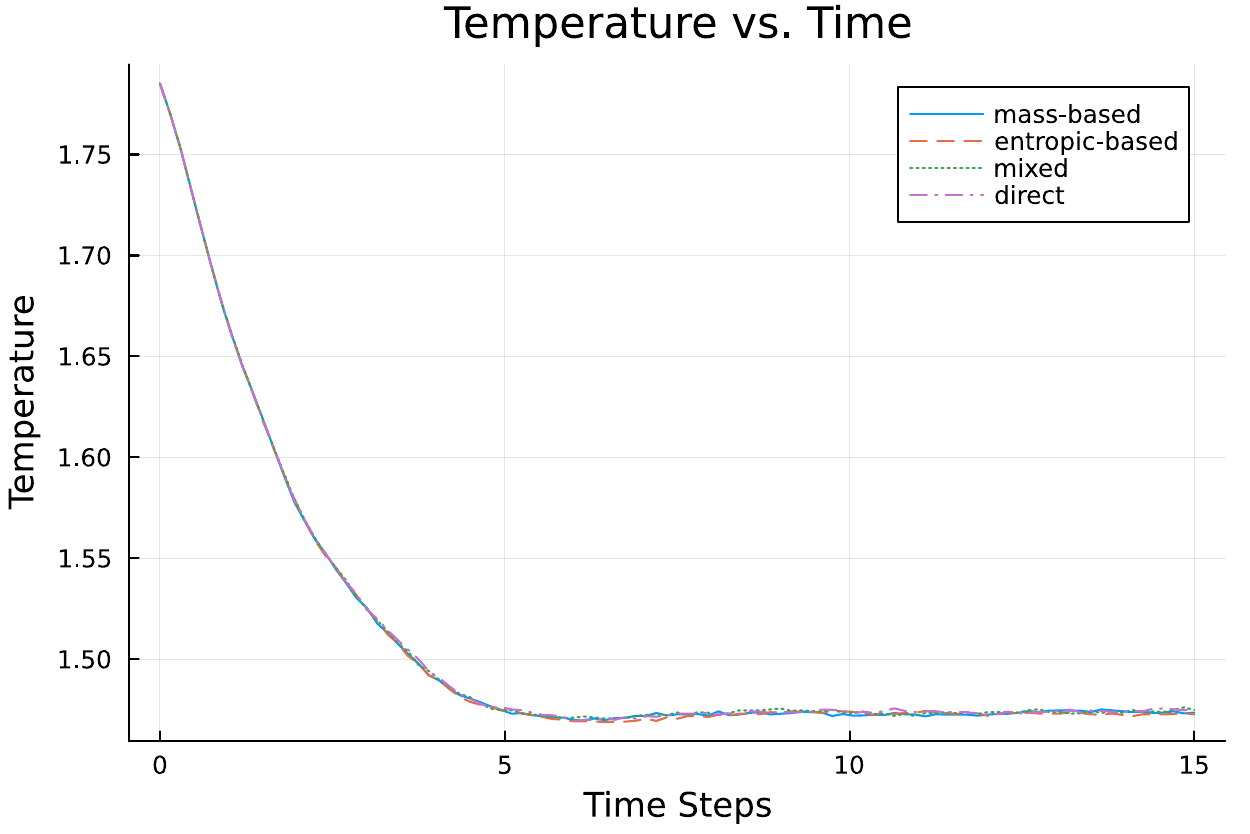}}
     {average temperature}
\end{tabular}
\caption{\label{fig.aE}The total energy error is of order $10^{-4}\%$ and decreases with the time step quadratically for four shown approaches: mass-based volume, entropic volume, direct volume, and mixed volume. The total temperature decreases and reaches the value given by laws of adiabatic expansion and all the four approaches give the same temperature profiles.}
\end{figure}


\section{Entropic SPH with dissipative evolution}\label{sec.dis}
In this Section, we show how to add dissipative processes to the entropic SPH (to all of its formulations). In particular, we include Fourier heat conduction and viscous dissipation, and finally we add algebraic dissipation in the context of hyperbolic heat conduction. 

It should be noted that dissipative terms have to update the particle entropies directly $S_\alpha$, since entropy densities $s_\alpha$ are calculated from the particle densities and not vice versa.

\subsection{Fourier heat conduction}\label{sec.fourier}
As we have entropy among state variables, we can capture the transformation of the kinetic energy into the internal energy (dissipation) while keeping the total energy constant. In other words, entropic SPH preserves the total energy also when dissipation (viscosity, heat conduction, etc.) is present.

Here we outline a model for classical Fourier heat conduction which is based on a dissipation potential with a gradient of conjugate state variable. The SPH evolution can be then seen as a realization of the General Equation for Non-Equilibrium Reversible-Irreversible Coupling (GENERIC) \cite{go,og, hco,pkg}. In order to discretize the dissipation potential, we need to discretize differential operators first, as is done in the standard SPH. 

Let us now recall the discrete gradient and divergence operators, 
\begin{subequations}
    \begin{align}
    \tilde{G}^0_\alpha(A_\square) =& -\sum_\beta V_\beta (A_\alpha-A_\beta) W'_{\alpha\beta}\ee_{\alpha\beta}\\
    D^0_\alpha(A_\square) =& \sum_\beta V_\beta (A_\alpha+A_\beta) W'_{\alpha\beta}\ee_{\alpha\beta},
    \end{align}
\end{subequations}
where $A_\square$ is a collection of discrete values $\{A_\alpha\}$ for all the SPH particles, see \cite{violeau}, and particle volume $V_\alpha$ is subjected to one of the choices above. Note that these operators satisfy the duality
\begin{equation}
    \langle A_\square, \tilde{G}^0_\alpha (B_\square)\rangle  = - 
    \langle D^0_\alpha(A_\square), B_\square\rangle,
\end{equation}
where $\langle A_\square, B_\square\rangle = \sum_\alpha V_\alpha A_\alpha B_\alpha$.

Then, the continuous dissipation potential 
\begin{equation}\label{eq.Xi.heat} 
\Xi^{(heat)} = \frac{1}{2}\int d\rr \lambda(T) (\nabla \epsilon^*)^2,
\end{equation}
where the conjugate internal energy $\epsilon^* = \frac{\delta S}{\delta \epsilon}$ can be interpreted as the inverse temperature $T^{-1}$,
can be discretized as
\begin{align}
	\Xi^{(SPH-heat)} =& \frac{1}{2}\sum_\alpha V_\alpha \lambda(T_\alpha) \tilde{G}^0_\alpha (\epsilon^*_\square) \tilde{G}^0_\alpha (\epsilon^*_\square)\nonumber\\
    =&\frac{1}{2}\sum_\alpha V_\alpha \lambda(T_\alpha) \sum_\beta V_\beta (\epsilon^*_\alpha - \epsilon^*_\beta)W'_{\alpha\beta}\ee_{\alpha\beta}\sum_\gamma V_\gamma(\epsilon^*_\alpha - \epsilon^*_\gamma)W'_{\alpha\gamma}\ee_{\alpha\gamma},
\end{align}
where $\epsilon^*_\alpha = T^{-1}_\alpha$ is the SPH inverse temperature of particle $\alpha$.  
Irreversible evolution of the discrete energy density $\epsilon_\alpha$ is then obtained as the functional derivative of the dissipation potential with respect to the conjugate variable $\epsilon^*$,\footnote{The $1/V_\alpha$ prefactor is present as energy \emph{density} is on the left hand side and the right hand side has to have the same property. In continuum thermodynamics, the functional derivative produces the density-like behavior.} which can be calculated from the variation of the dissipation potential,
\begin{align}
	\Xi^{(SPH-heat)}(\epsilon^*+\delta\epsilon^*) =& \frac{1}{2}\sum_\alpha V_\alpha \lambda(T_\alpha) \tilde{G}^0_\alpha (\epsilon^*) \tilde{G}^0_\alpha (\epsilon^*)
	+ \sum_\alpha V_\alpha \lambda(T_\alpha) \tilde{G}^0_\alpha (\delta \epsilon^*) \tilde{G}^0_\alpha ( \epsilon^*) + \mathcal{O}(\delta\epsilon^*)^2\nonumber\\
	=& \Xi^{(SPH-heat)}(\epsilon^*) 
	-\sum_\alpha V_\alpha D^0_\alpha(\lambda(T_\square) \tilde{G}^0_\square(\epsilon^*_\square))\delta \epsilon^*_\alpha + \mathcal{O}(\delta\epsilon^*)^2,
\end{align}
as
\begin{equation}\label{eq.ea.gd}
    \dot{\epsilon}_\alpha = \frac{1}{V_\alpha}\frac{\partial \Xi}{\partial \epsilon^*_\alpha} = - D^0_\alpha (\qq_\square) = -\sum_\beta V_\beta (\qq_\alpha+\qq_\beta)\cdot W'_{\alpha\beta}\ee_{\alpha \beta} 
\end{equation}
where
\begin{equation}\label{eq.q}
	\qq_\alpha = \lambda(T_\alpha) \tilde{G}^0_\alpha (\epsilon^*_\square) = -\sum_\beta V_\beta\lambda(T_\alpha) (\epsilon^*_\alpha - \epsilon^*_\beta) W'_{\alpha\beta}\ee_{\alpha\beta}
	= -\sum_\beta V_\beta\lambda_F (T_\alpha - T^2_\alpha/T_\beta) W'_{\alpha\beta}\ee_{\alpha\beta}
\end{equation} 
is the discrete heat flux, $\lambda(T_\alpha)=\lambda_F T_\alpha^2$ for $\lambda_F$ being the Fourier heat conductivity, and $s_\alpha$ being the particle entropy density. The result is then similar to the usual approach towards heat conduction in SPH \cite{monaghan-review,jeong2003}, except for replacing $T_\alpha-T_\beta$ with $T_\alpha-T_\alpha^2/T_\beta$.

Discrete heat equation \eqref{eq.ea.gd} satisfies the second law of thermodynamics, which can be shown as 
\begin{equation}
    \dot{S} = \sum_\alpha V_\alpha \epsilon^*_\alpha \dot{\epsilon}_\alpha 
    = - \langle \epsilon^*_\alpha, D^0_\alpha \qq \rangle  
    = \langle \tilde{G}^0_\alpha \epsilon^*, \qq_\alpha\rangle 
    = \left\langle \frac{\qq_\alpha}{\lambda(T_\alpha)}, \qq_\alpha \right\rangle\geq 0.
\end{equation}
However, this could already be inferred from the convexity of the continuous dissipation potential \eqref{eq.Xi.heat} because the dissipation potential is convex \cite{jsp2020} and composition of a convex and affine function is convex as well\footnote{The second derivative of composition $\Xi\circ f(y)$, where $f$ is affine, is $\frac{\partial^2 \Xi}{\partial f^2}\left(\frac{\partial f}{\partial y}\right)^2\geq 0$.}.

\paragraph{Mass-based volume approach.}
In this case, we consider the mass-based volume, that is, $V_{\alpha}=V_{\alpha}^{m}=\frac{m_\alpha}{\rho_\alpha}$. Then, as mentioned above, the dissipative processes are included via an update in the discrete Lagrangian quantities. In the case of Fourier heat conduction, we need to update the entropy $S_\alpha$. This is done by  taking
\begin{equation}
    \dot{S}_\alpha = V_{\alpha}^{m}\frac{\partial s_\alpha}{\partial \epsilon_\alpha} \dot{\epsilon}_\alpha = -\epsilon^*_\alpha \sum_\beta V_\alpha^{m} V_\beta^{m} (\qq_\alpha+\qq_\beta)\cdot W'_{\alpha\beta}\ee_{\alpha \beta},
\end{equation}
where $\qq_\alpha$ is given by Equation \eqref{eq.q} with $V_\alpha = V^m_\alpha$. This dissipative evolution is then added to the Hamiltonian evolution equations according to the GENERIC framework.


\paragraph{Entropic-based volume and mixed-based volume approach.}
As in the mass-based approach, we need to convert the Eulerian change of entropy $\dot{s}_\alpha$ to the Lagrangian one, $\dot{S}_\alpha$. This is achieved again by taking
\begin{equation}
    \dot{S}_\alpha = V_{\alpha}^{s}\epsilon^*_\alpha \dot{\epsilon}_\alpha 
    = -\epsilon^*_\alpha \sum_\beta V^s_\alpha V^s_\beta (\qq_\alpha+\qq_\beta)\cdot W'_{\alpha\beta}\ee_{\alpha \beta},
\end{equation}
where $\qq_\alpha$ is given by Equation \eqref{eq.q}  and $V_{\alpha}=V_{\alpha}^{s}=\frac{S_\alpha}{s_\alpha}$. The same relation is also used in the mixed-volume-based approach.


\paragraph{Implicit-based volume approach}
We again need to provide a way to update the Lagrangian particle entropy $S_{\alpha}$ due to dissipation while respecting the definition of the implicit volume $V_{\alpha}^{I}$. Taking the advantage of the assumption made earlier that the dissipative dynamics does not alter particle positions, 
that is $\frac{\delta S_{\alpha}}{\delta x_{\beta}}=0$, differential of the definition of the entropy-based volume gives
\begin{equation}
    0 = \sum_\beta \left(\frac{d S_\beta}{s_\beta}-\frac{S_\beta}{s^2_\beta}ds_\beta\right)W_{\alpha\beta}.
\end{equation}
This is satisfied when the particle entropy is updated as
\begin{equation}
    \dot{S}_\alpha = \frac{S_\alpha}{s_\alpha}\dot{s}_\alpha.
  \end{equation}
  Analogical update rules for the Lagrangian entropies $S_\alpha$ are used also in the presence of other dissipation sources, as for instance the viscous dissipation in the following Section.

\subsection{Viscous dissipation} 
Although viscous terms can also be formulated as gradient dynamics on the continuous level, see \cite{pkg}, its discretization is not as straightforward as in the case of heat conduction. The problem lies in the Lagrangian nature of the momentum in SPH, in contrast to the Eulerian character of the entropy $s_\alpha$ (or density $\rho_\alpha$). In other words, an SPH discretization of the continuous dissipation potential \cite{pkg} would produce a discrete evolution for the Eulerian momentum density $\mm_\alpha = \sum_\beta W_{\alpha\beta}M_\beta$, not for $\MM_\alpha$ itself. Therefore, we opt out of adapting a usual SPH discretization of the viscous terms \cite{monaghan-viscous},
\begin{equation}\label{eq.vis.SPH}
(\dot\MM_\alpha)_{diss} = \sum_\beta 2(n+2)\mu V_{\alpha} V_{\beta}(\vv_{\alpha}-\vv_{\beta})\cdot\ee_{\alpha\beta}\frac{W'_{\alpha\beta}}{r_{\alpha\beta}} \ee_{\alpha\beta},
\end{equation}
where $n$ is the geometric dimension of the system.

Viscous dissipation also changes the kinetic energy of particle $\alpha$ as
\begin{equation}
  \dot{e}^{kin}_{\alpha}= \frac{d}{dt}\left( \frac{1}{2} \vv_{\alpha}\MM_{\alpha}\right) = \vv_{\alpha}\dot{\MM}_{\alpha} =  2(n+2)\mu\vv_\alpha\cdot \sum_\beta \ee_{\alpha\beta} V_{\alpha}V_{\beta} (\vv_{\alpha}-\vv_{\beta})\cdot\ee_{\alpha\beta}\frac{W'_{\alpha\beta}}{r_{\alpha\beta}},
\end{equation}
which eventually reduces the total kinetic energy 
\begin{equation}
\dot{e}^{kin} = 
\sum_\alpha \dot{e}^{kin}_{\alpha} =  (n+2)\mu\sum_\alpha \sum_\beta ((\vv_{\alpha}-\vv_{\beta})\cdot\ee_{\alpha\beta})^2 V_{\alpha} V_{\beta} \frac{W'_{\alpha\beta}}{r_{\alpha\beta}} \leq 0.
\end{equation}
The sign of this inequality is caused by the assumed monotonicity of the $W(r)$ smoothing kernel.

While the kinetic energy is reduced, the internal energy grows so that the total energy is conserved. Internal energy is a function of mass density and entropy density, and the growth of internal energy is caused by the entropy production
\begin{equation} \label{eq.SupdateViscous}
(\dot{S}_\alpha)_{diss} = -\frac{1}{T_\alpha} \sum_\beta (n+2)\mu V_{\alpha}V_{\beta}((\vv_{\alpha}-\vv_{\beta})\cdot\ee_{\alpha\beta})^2 \frac{W'_{\alpha\beta}}{r_{\alpha\beta}}\geq 0.
\end{equation}
Then the overall entropy $S = \sum_\alpha S_\alpha$ is produced, $\dot{S}\geq 0$, and the second law of thermodynamics is satisfied. The overall energy $E = E^{kin} + E^{int}$ is not changed, at least as far as the continuous form of the entropic SPH, before the temporal discretization, is considered. Note that, as a consequence, we shall use Equation \ref{eq.SupdateViscous} as an update for the Lagrangian particle entropy $S_{\alpha}$

\subsection{Illustration  - non-adiabatic expansion} 
\begin{figure}[ht!]
\begin{center}
\centering
\begin{tabular}{cc}
\subf{\includegraphics[width=60mm]{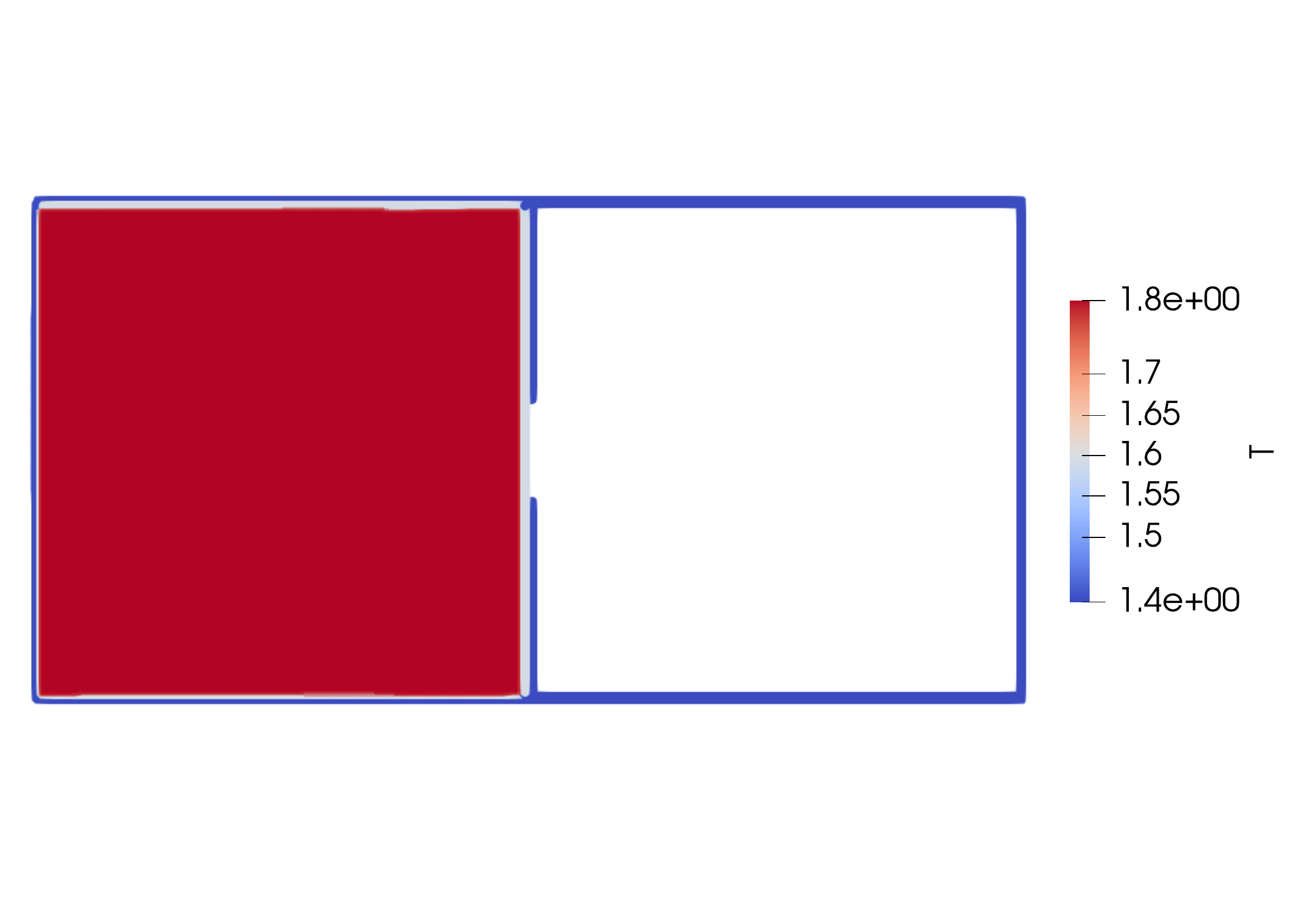}}
     {$t=0.0$}
&
\subf{\includegraphics[width=60mm]{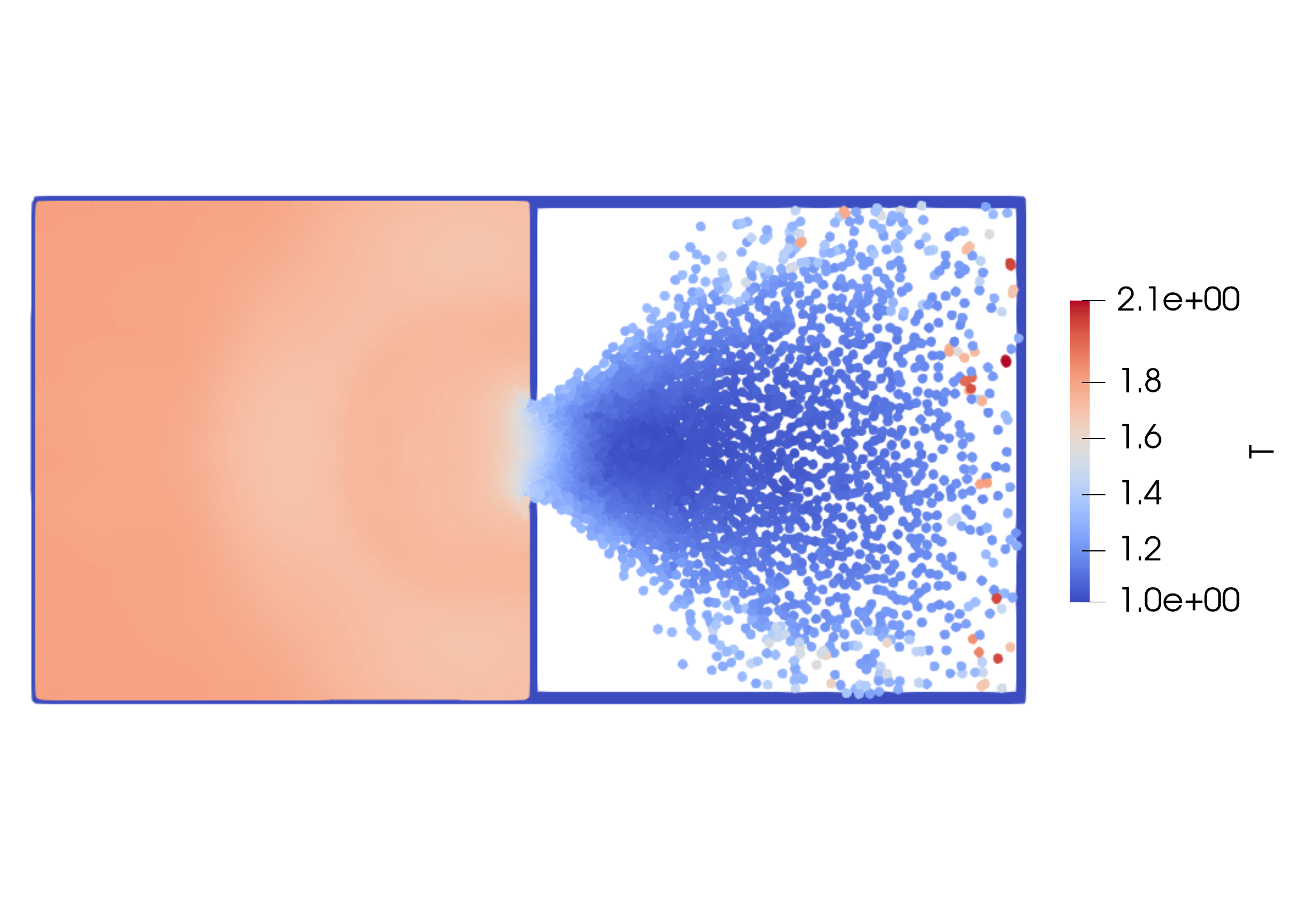}}
     {$t=0.6$}
\\
\subf{\includegraphics[width=60mm]{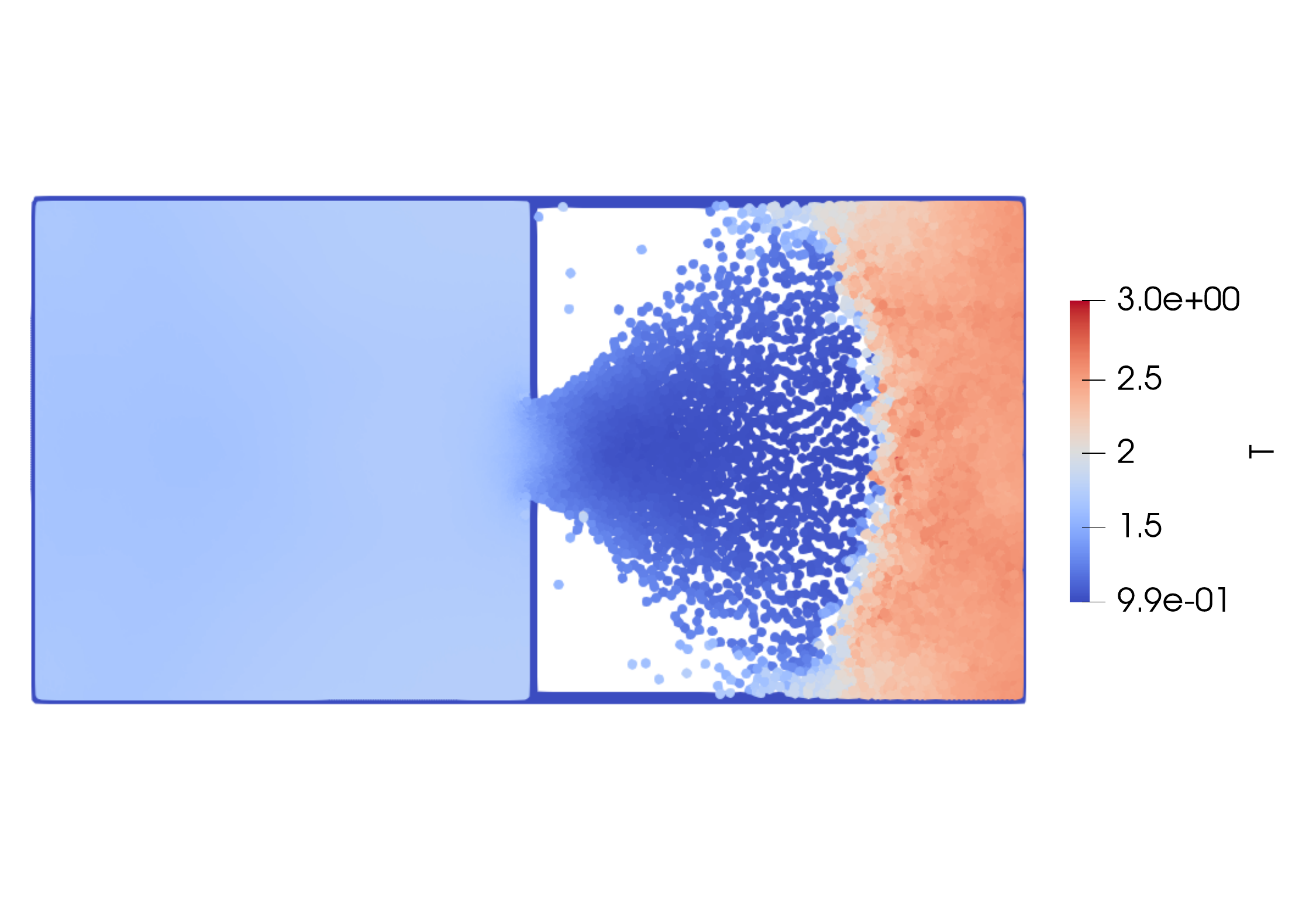}}
     {$t=1.5$}
&
\subf{\includegraphics[width=60mm]{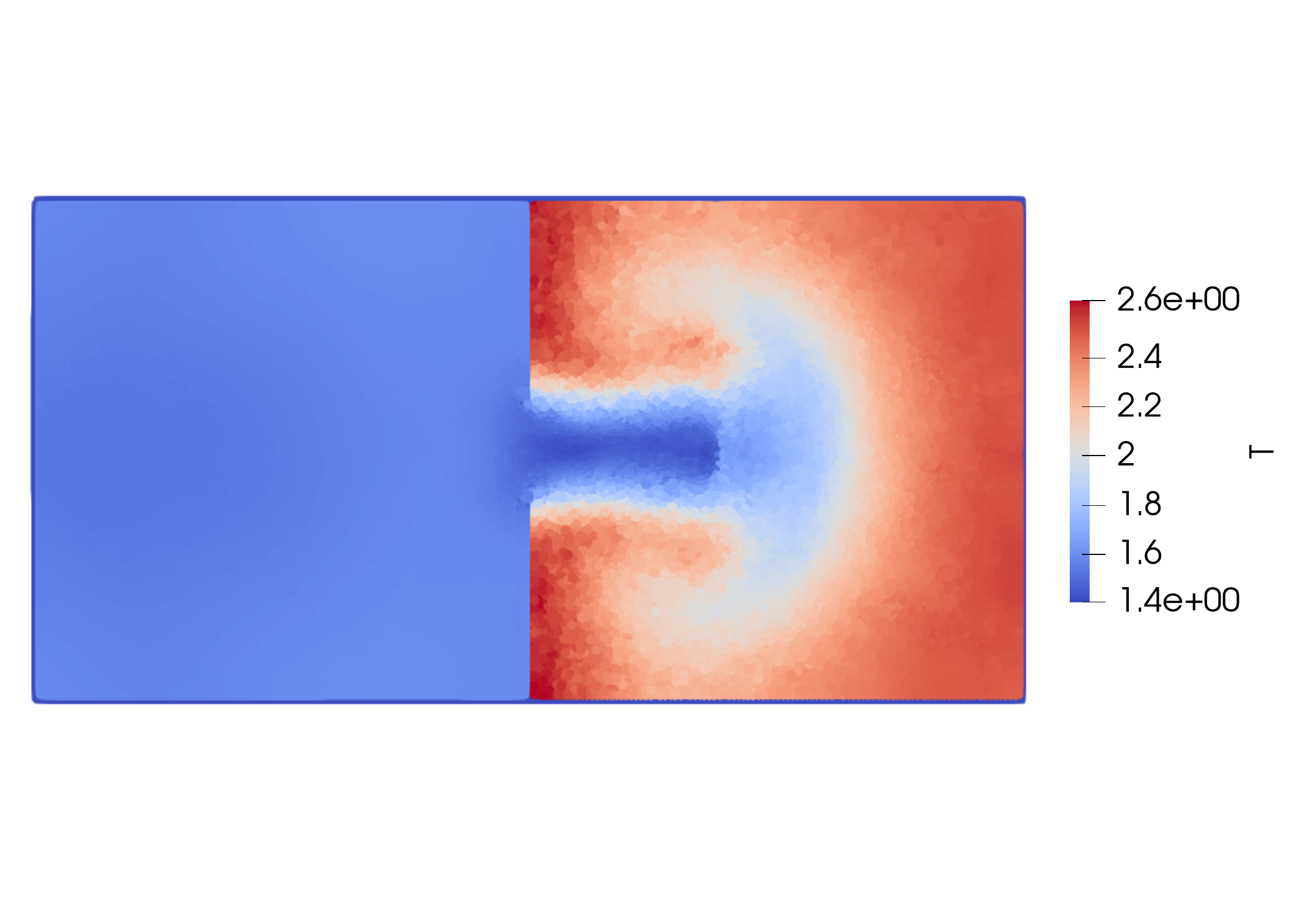}}
     {$t=3.5$}
\end{tabular}
\caption{\label{fig.nonadiabatic}Non-adiabatic expansion (43017 particles). All particles are initially in the left compartment only and gradually occupy the whole volume. Their temperature drops as in an adiabatic process, but then increases due to viscous ($\mu=0.001$) and Fourier dissipation ($\lambda = 0.1$). The heat flux was pointing to the left almost everywhere in the domain. The simulation was carried out with the mass-volume approach, and the entropic and mixed-volume approaches give quite similar results.}
\end{center}
\end{figure}

Let us now illustrate the Fourier and viscous dissipative effects on a non-adiabatic expansion, where the dissipative effects raise the entropy. Figure \ref{fig.nonadiabatic} shows the results for the mass-volume method (the entropic and mixed methods give practically the same results). Instead of a temperature drop as in the adiabatic case, we observe that the temperature returns to its initial value because of the dissipative effects.

\begin{figure}[ht!]
\centering
\begin{tabular}{cc}
\subf{\includegraphics[width=60mm]{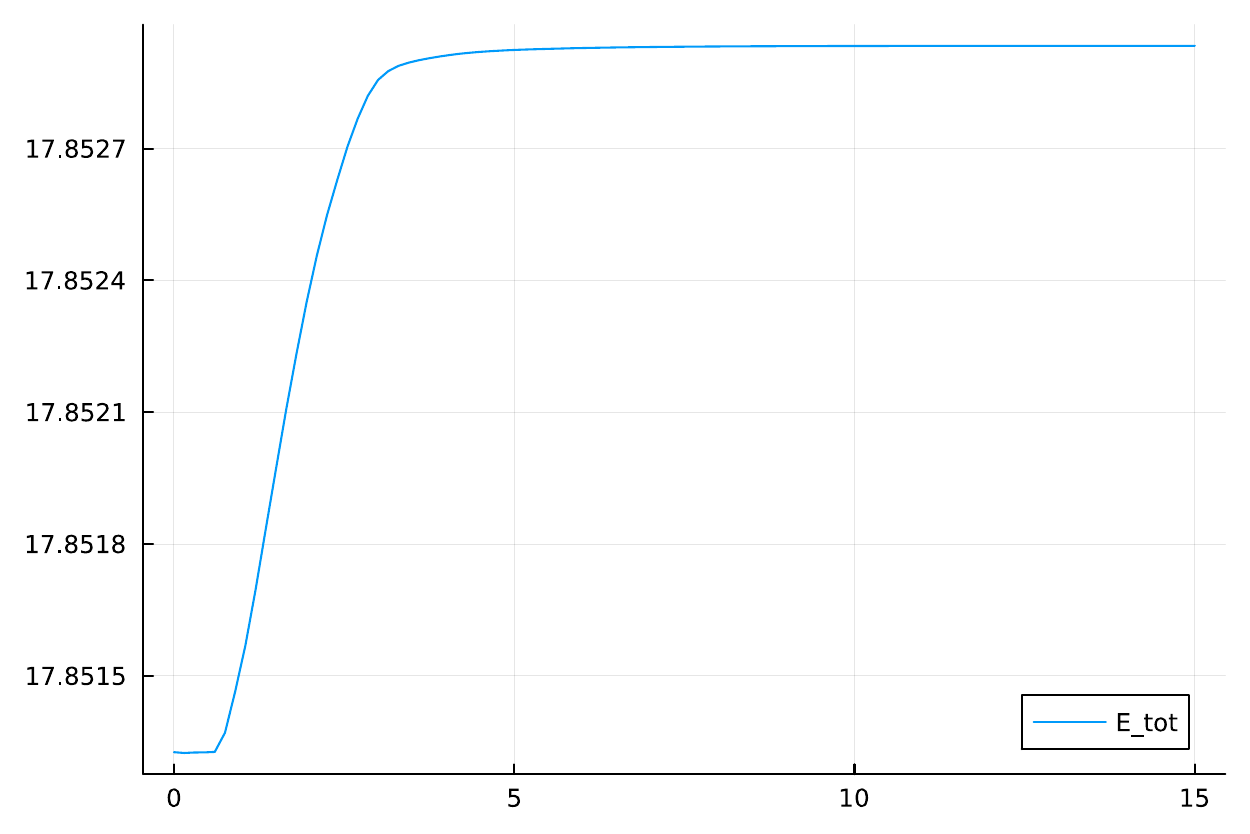}}
     {total energy in time}
&
\subf{\includegraphics[width=60mm]{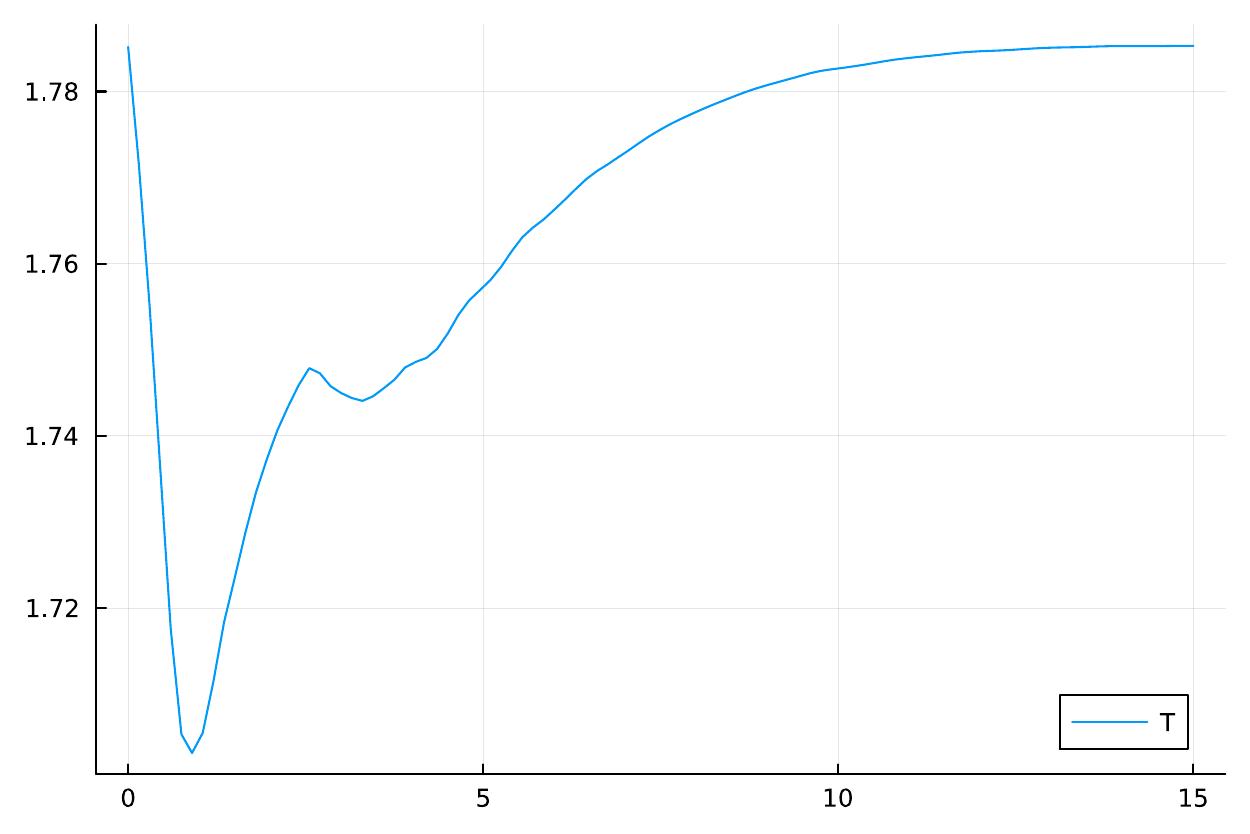}}
	{average temperature}
\end{tabular}
\caption{Total energy in time gradually increases due to numerical errors in the dissipation (by 0.006\%). However, the error decreases with the time step (linearly, as it is dominated by the first-order dissipative terms). The temperature initially drops due to the nearly adiabatic expansion and then recovers its initial value due to the dissipative effects.\label{fig.avE}}
\end{figure}

Figure \ref{fig.avE} shows that the energy increases slightly due to the numerical errors in the first-order time discretization of the dissipative terms. The error decreases with the time step and could be further decreased, for instance, by the Runge-Kutta method \cite{runge,kutta,kincl-unified}.

\subsection{Illustration - Rayleigh-Bénard convection}
Our final example is a vessel with a fluid described by the stiffened gas equation of state (see Appendix \ref{sec.thermo} or \cite{ader-vis}) so that it has only limited but non-zero compressibility. The fluid is in a gravitational field and is subject to heating from the bottom boundary as in the Rayleigh-Bénard convection \cite{benard,rayleigh-benard,lebon-understanding}. Note that we do not use the usual Boussinesq approximation \cite{boussinesq} as the fluid is treated as compressible. We use the mass-based volume approach (11k particles, viscosity $\mu=8.4\cdot 10^{-4}$, and heat conductivity $\lambda_F = 1.0\cdot 10^{-4}$).

\begin{figure}[ht!]
\centering
\begin{tabular}{cc}
\subf{\includegraphics[width=80mm]{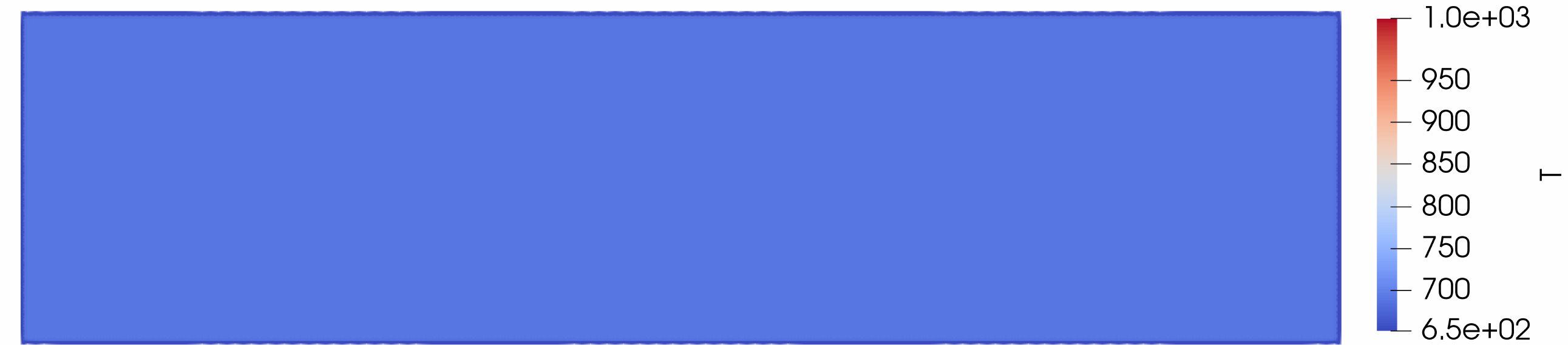}}
     {$t=0.0$}
&
\subf{\includegraphics[width=80mm]{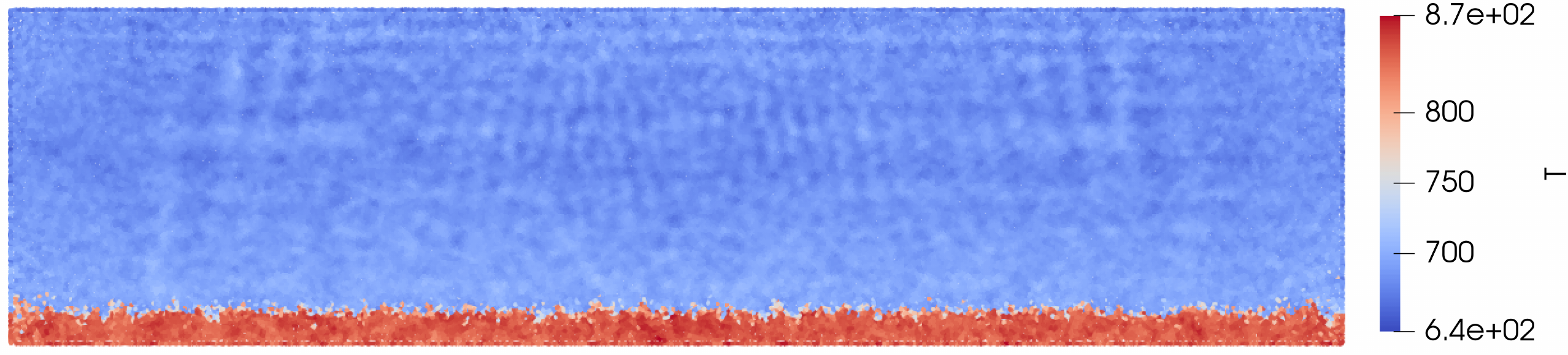}}
     {$t=0.003$}
\\
\subf{\includegraphics[width=80mm]{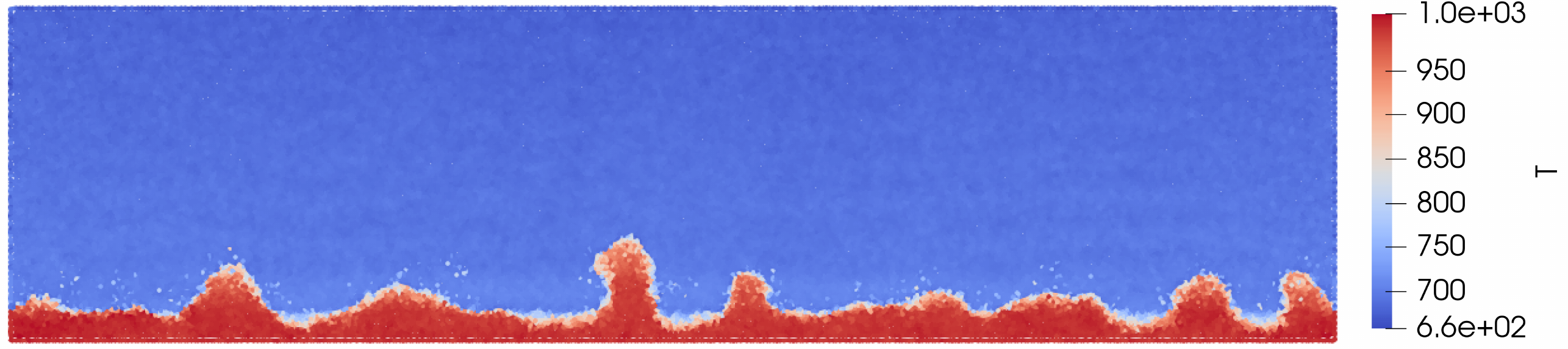}}
     {$t=0.006$}
&
\subf{\includegraphics[width=80mm]{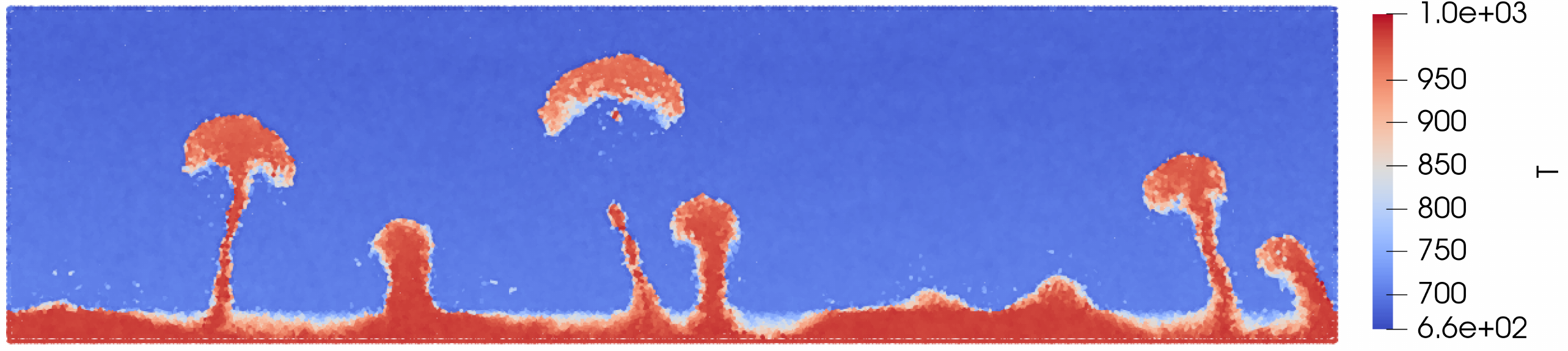}}
     {$t=0.05$}
\end{tabular}
\caption{\label{fig.heat}Temperature in a fluid with heated bottom boundary and cooled upper boundary. The first left-top figure shows the initial condition, the right-top figure shows the initial hydrostatic equilibrization, the left-bottom figure shows the onset of instability, and the right-bottom figure contains the mushroom-like plumes. The instability is caused by buoyancy due to the lower density of the fluid near the bottom boundary, leading to the formation of mushroom-like plumes. Both viscous dissipation and Fourier heat conduction were switched on.}
\end{figure}

After the initial equilibration of the gravitational energy, we observe mushroom-like structures caused by the buoyancy due to lower density the heated parts of the fluid, see Figure \ref{fig.heat}.


  \subsection{Hyperbolic heat conduction -- Lagrangian approach}\label{sec.hyp}
Although classical Fourier heat conduction is usually sufficient for the description of heat transfer, very fast heat pulses require hyperbolic heat conduction \cite{naf,kovacs-Van}. Fourier heat conduction can be seen to be an approximation of hyperbolic heat conduction for sufficiently long times \cite{jou-eit}. There are two routes towards evolution equations for hyperbolic heat conduction. The first route starts with the kinetic theory of phonons (quasiparticles present in a crystal lattice) \cite{peiersl}. The kinetic theory can be reduced to a less detailed theory, taking into account only a few moments of the distribution function \cite{struch-dreyer}, or the Eulerian entropy density and phonon momentum density. This route leads to the Eulerian hyperbolic heat conduction studied in the subsequent Section. 

In this Section, we follow the second route that leads to a Lagrangian version of hyperbolic heat conduction. We commence by introducing a scalar field $\psi(\XX)$ canonically coupled to the field of Lagrangian entropy density, $s_0(\XX)$, via Poisson bracket
\begin{equation}
\{F,G\}^{(s_0-\psi)} = \int d\XX \left( F_{s_0}G_{\psi} - G_{s_0}F_{\psi}\right).
\end{equation}
Instead of the scalar field $\psi$, only its gradient $W_I=\frac{\partial \psi}{\partial X^I}$ is considered, which will represent a vectorial quantity related to the heat flux \cite{shtc-generic,God-Siberian,Malyshev-Romenski}. Finally, Lagrangian quantity $W_I$ is projected to its Eulerian counterpart 
\begin{equation}\label{eq.w}
w_i = \frac{\partial X^I}{\partial x^i}W_I = \frac{\partial \psi(\XX(\xx))}{\partial x^i},
\end{equation}
called the conjugate entropy flux, as the derivative of energy with respect to $\ww$ turns out to be the entropy flux.
In SPH, we shall follow the second route, since it leads to similar results as the first route (based on kinetic theory \cite{fononi}) while being more suitable for the SPH discretization.

Within SPH, hyperbolic heat conduction has already been implemented in \cite{jiang2006} by discretization of the resulting Cattaneo equation (a telegraphist's equation for temperature). Here, we formulate another model within SPH that is Hamiltonian and symplectic (when disregarding the dissipation) and thus suitable for symplectic integrators. 
Based on the above geometric construction of the continuous hyperbolic heat conduction, we introduce the SPH variables for hyperbolic heat conduction as
\begin{subequations}
\begin{align}
S_\alpha =& \int d\XX \chi_\alpha(\XX) s_0(\XX)\\
\ww_\alpha =& \tilde{G}_\alpha^0(\psi_\square)
\end{align}
    where $\psi_\alpha = \int d\XX \bar{\chi}_\alpha(\XX) \psi(\XX)$.
\end{subequations}
The Poisson bracket for variables $S_\alpha$ (entropy of particle $\alpha$) and $\ww_\alpha$ (conjugate entropy flux of particle $\alpha$) is calculated from the canonical Poisson bracket for $s_0$ and $\psi$, 
\begin{equation}\label{eq.PB.Sa-wa}
\{F,G\}^{(S_\alpha, \ww_\alpha)} = -\sum_\alpha V_\alpha F_{S_{\alpha}} D_\alpha^0\left(\frac{G_{\ww_\square}}{V_\square}\right)-\sum_\alpha \tilde{G}_\alpha^0(G_{S_\square}) F_{\ww_\alpha}.
\end{equation}
The reversible evolution equations implied by this bracket are
\begin{subequations}\label{eq.Sw.rev}
\begin{align}
(\dot{S}_\alpha)_{rev} =& -\sum_\beta V_\alpha V_\beta \left(\frac{E_{\ww_\alpha}}{V_\alpha} + \frac{E_{\ww_\beta}}{V_\beta}\right)\cdot W'_{\alpha\beta}\ee_{\alpha\beta}\\
(\dot{\ww}_\alpha)_{rev} =& \sum_\beta V_\beta (E_{S_\alpha} - E_{S_\beta}) W'_{\alpha\beta}\ee_{\alpha\beta}.
\end{align}
\end{subequations}
These equations automatically conserve the total entropy, $\dot{S}=\sum_\alpha \dot{S}_\alpha = 0$, as well as the total energy $E(S_\square,\ww_\square)$. It can be seen from the equation for entropy that the derivative of energy with respect to the $\ww$ field is the entropy flux, and thus $\ww_\alpha$ is called the conjugate entropy flux at the particle $\alpha$.

Apart from the reversible evolution, the conjugate entropy flux $\ww_\alpha$ has also irreversible evolution that drives it towards an equilibrium value, 
\begin{subequations}\label{eq.Sw.irr}
\begin{align}
(\dot{\ww}_\alpha)_{irr} = -\frac{1}{\tau} E_{\ww_\alpha}.
\end{align}
This dissipative evolution, which represents collisions of phonons with crystal impurities \cite{calaway, struch-dreyer}, is accompanied with entropy production
\begin{equation}
(\dot{S}_\alpha)_{irr} = \frac{1}{T_\alpha \tau} (E_{\ww_\alpha})^2 \geq 0,
\end{equation}
so that the total energy is conserved. Note that $\tau > 0$ is a relaxation parameter that characterizes collisions.
\end{subequations}

Equations \eqref{eq.Sw.rev} and \eqref{eq.Sw.irr} are then summed to the final evolution equations for the SPH hyperbolic heat conduction,
\begin{subequations}\label{eq.Sw}
\begin{align}
\dot{S}_\alpha =& -\sum_\beta V_\alpha V_\beta \left(\frac{E_{\ww_\alpha}}{V_\alpha} + \frac{E_{\ww_\beta}}{V_\beta}\right)\cdot W'_{\alpha\beta}\ee_{\alpha\beta}+\frac{1}{T_\alpha \tau} (E_{\ww_\alpha})^2\\
\dot{\ww}_\alpha =& \sum_\beta V_\beta (E_{S_\alpha} - E_{S_\beta}) W'_{\alpha\beta}\ee_{\alpha\beta}-\frac{1}{\tau} E_{\ww_\alpha}.
\end{align}
\end{subequations}

\begin{figure}[ht!]
\centering
\begin{tabular}{cc}
\subf{\includegraphics[width=60mm]{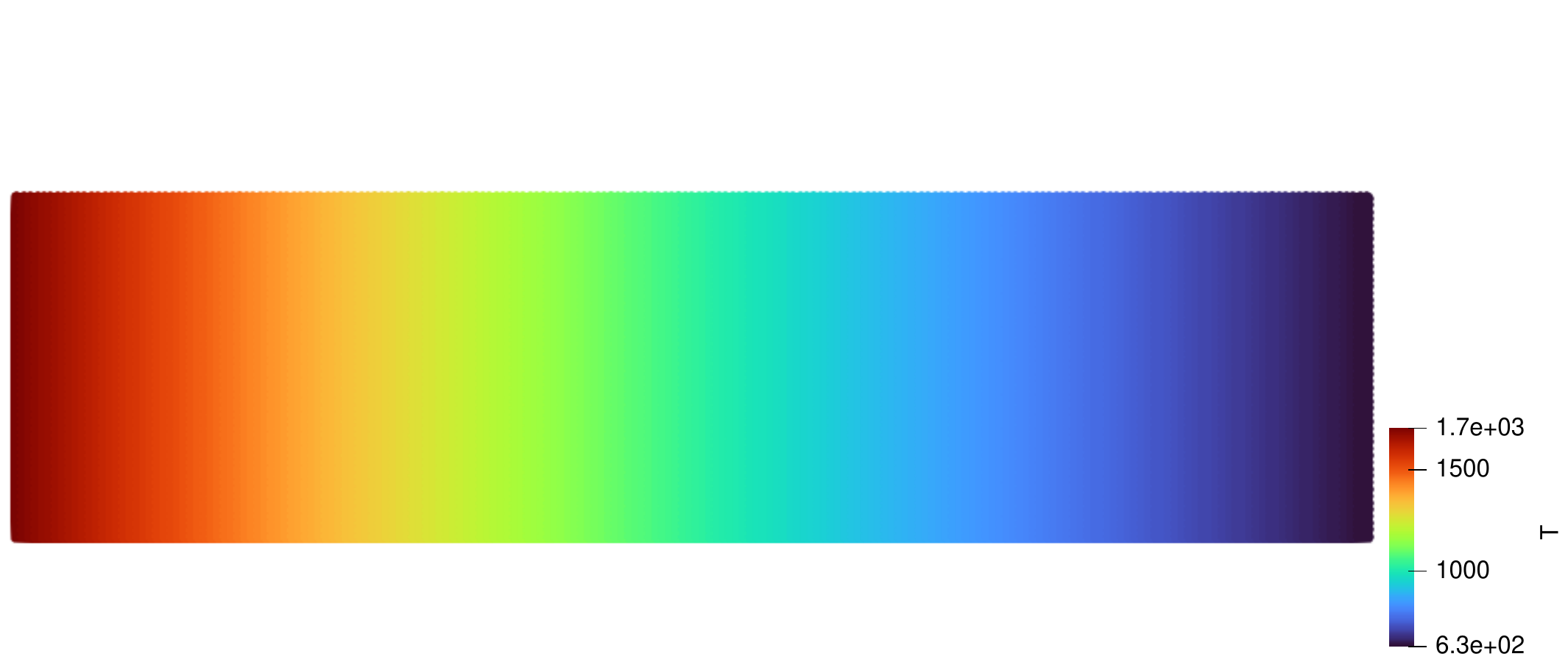}}
     {$t=0.0$}
&
\subf{\includegraphics[width=60mm]{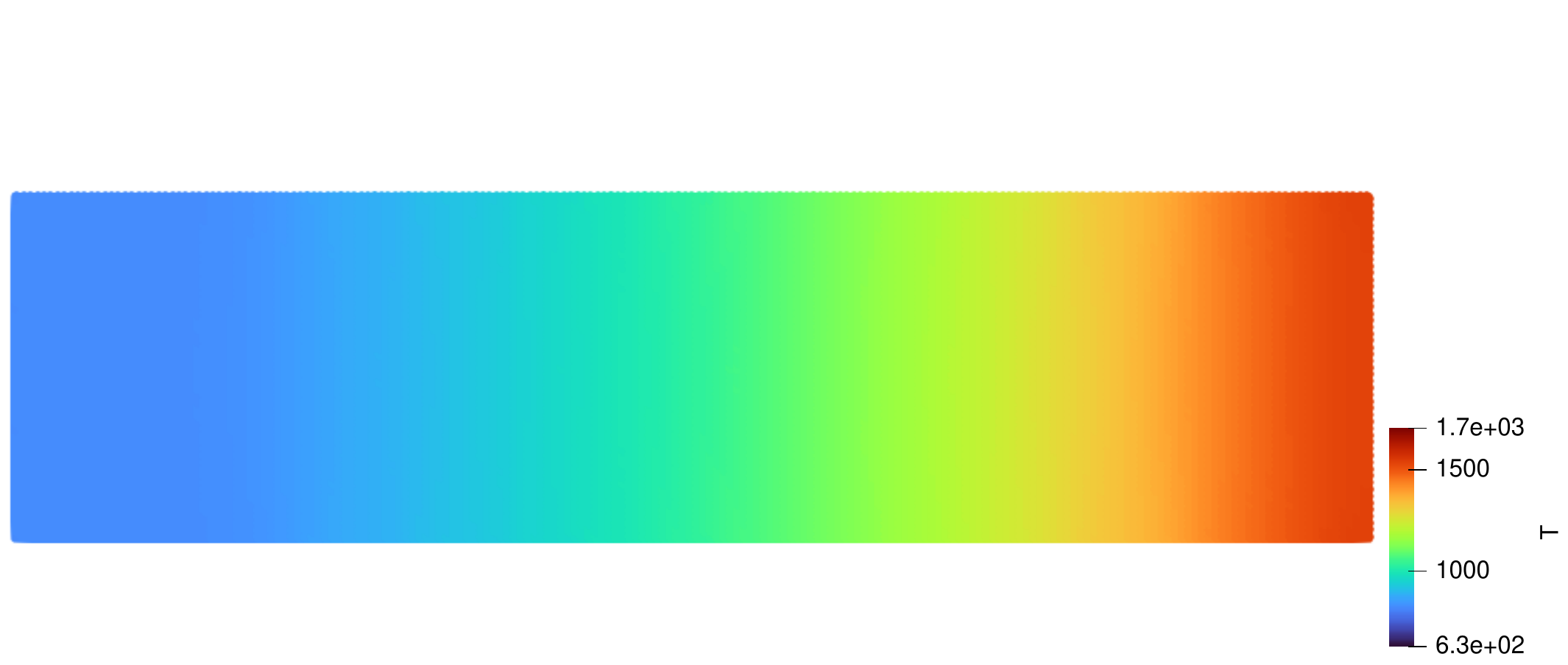}}
     {$t=0.59$}
\\
\subf{\includegraphics[width=60mm]{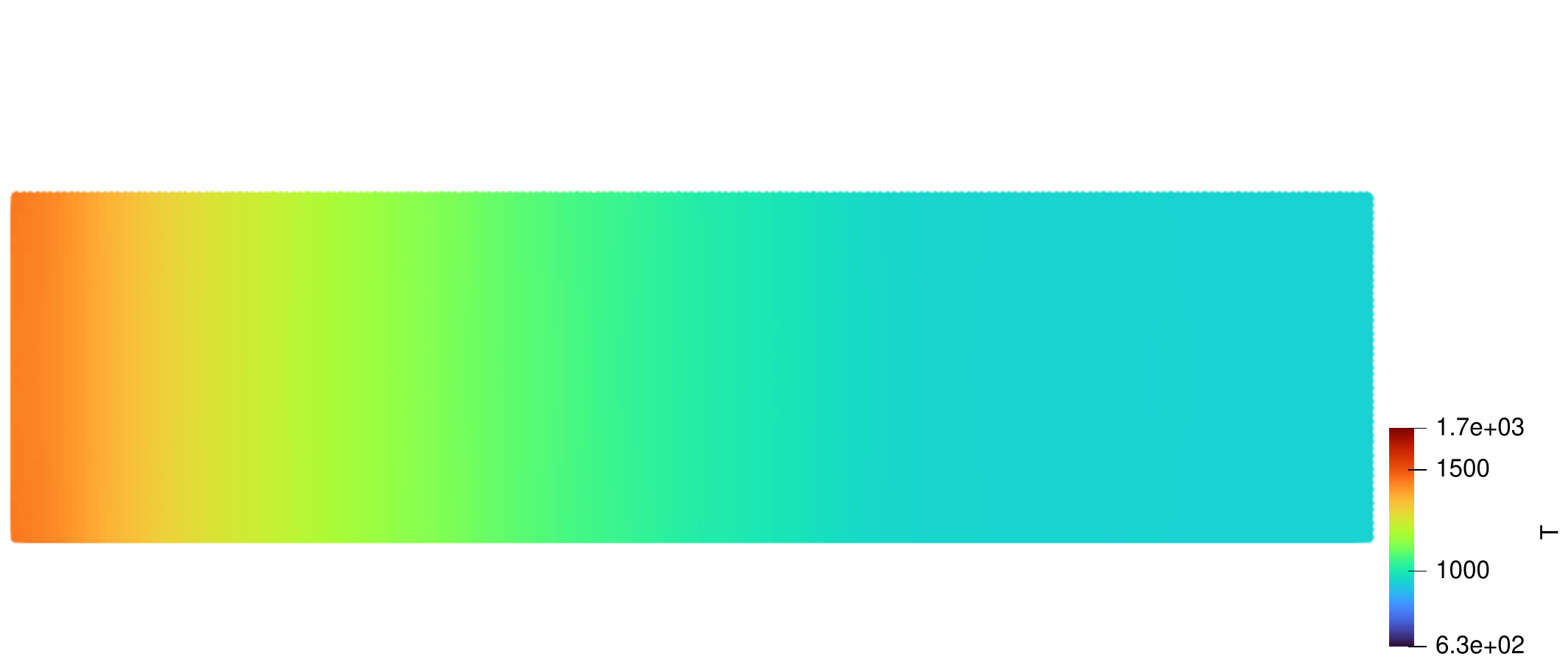}}
     {$t=1.14$}
&
\subf{\includegraphics[width=60mm]{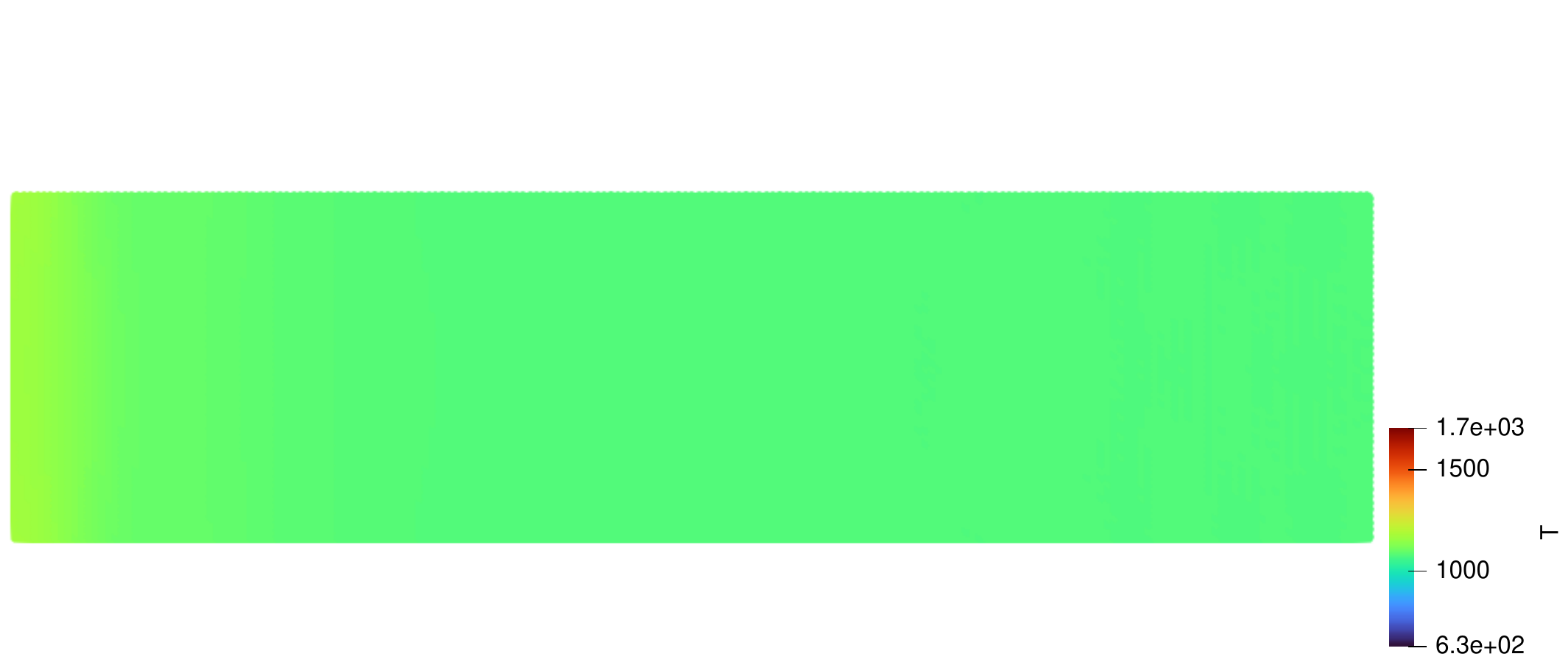}}
     {$t=4.69$}
\end{tabular}
    \caption{\label{fig.cattaneo} Four snapshots of the rectangular domain with hyperbolic heat propagation. While the heat waves travel through the domain and reflect, they are also being damped by dissipation. The energy was chosen as the sum of the kinetic energy of the heat flux $\sum_\alpha 0.05\frac{m_\alpha}{\rho_\alpha}\ww_\alpha^2$ and the internal energy of the stiffened gas ($c_V=1.0$, $p_0=10$, $\rho_0=10$, $\gamma=1.6$), see Appendix \ref{sec.stiffened}. The relaxation parameter was chosen as $\tau =10^{-5}$.}
\end{figure}

\begin{figure}[ht!]
\centering
\begin{tabular}{cc}
\subf{\includegraphics[width=60mm]{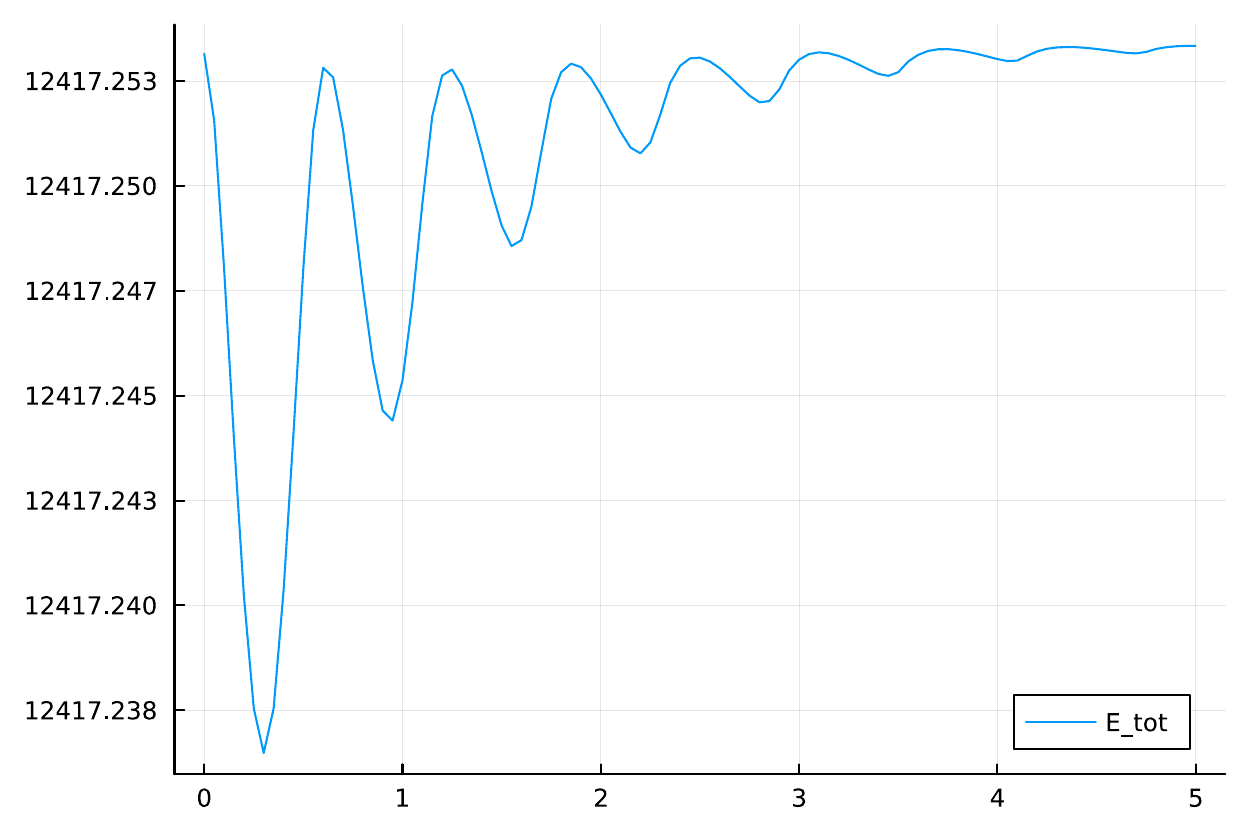}}
     {total energy in time}
&
\subf{\includegraphics[width=60mm]{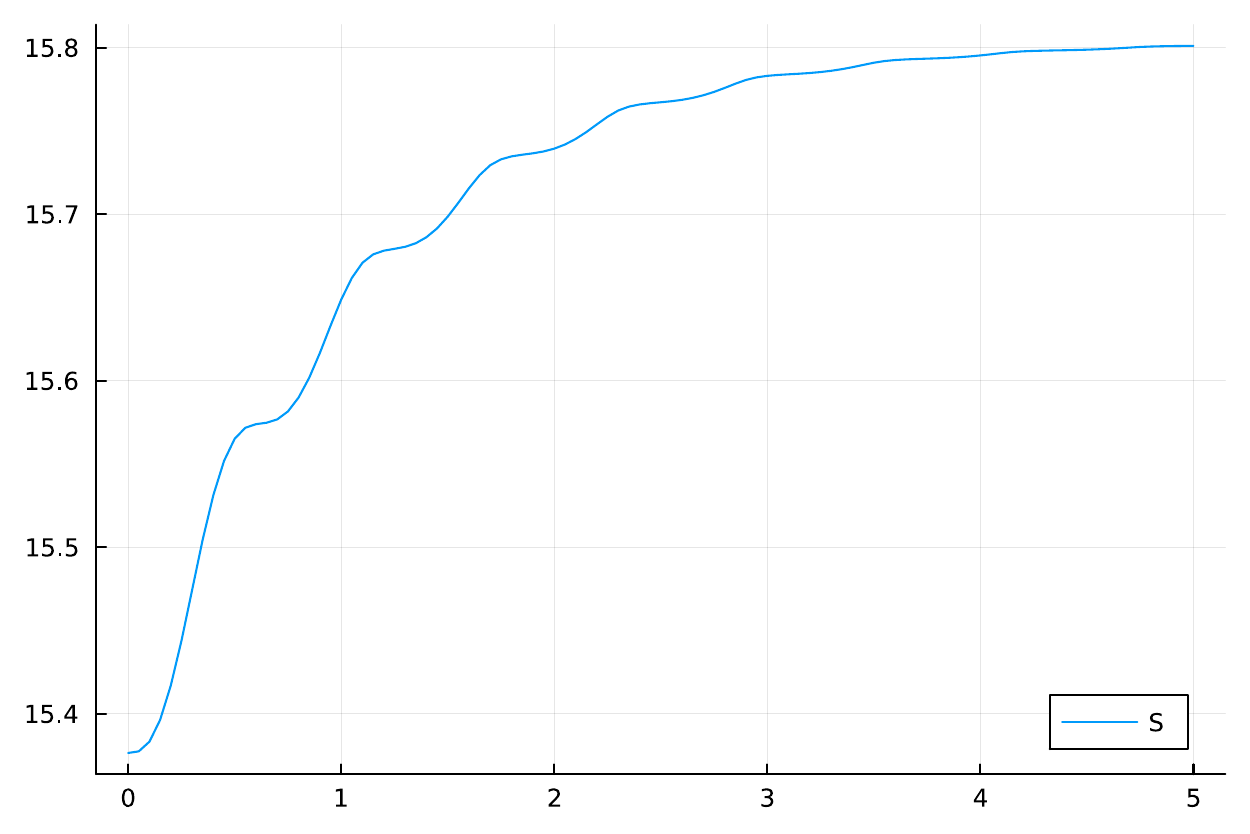}}
     {total entropy in time}
\end{tabular}
\caption{While the total energy is approximately conserved, the overall entropy is raised as expected \cite{jou-eit}.
\label{fig.cE}}
\end{figure}

Figure \ref{fig.cattaneo} shows four snapshots of the temperature in a rectangular domain with an initial horizontal temperature gradient. Figure \ref{fig.cE} shows that the total energy error was less than $10^{-4}\%$ and that the second law of thermodynamics was satisfied (entropy growing).

\textbf{Reduction to Fourier heat conduction}. Finally, Equations \eqref{eq.Sw} can be reduced to an equation for $S_\alpha$ when assuming that the equation for $\ww_\alpha$ relaxes quickly to its equilibrium, $\dot{\ww}_\alpha\approx 0$,
\begin{equation}\label{eq.wrelax}
    \frac{\qq_\alpha}{T_\alpha} = E_{\ww_\alpha} = \tau\sum_\beta V_\beta (E_{S_\alpha} - E_{S_\beta}) W'_{\alpha\beta}\ee_{\alpha\beta} = -\tau \tilde{G}_\alpha^0 (T_\square).
\end{equation}
Note that the entropy flux is equal to the heat flux divided by temperature, $E_{\ww_\alpha} =\qq_\alpha/T_\alpha$.
When we plug this value of $E_{\ww_\alpha}$ into the equation for $S_\alpha$ (as in the Dynamic MaxEnt method \cite{dynmaxent}), we obtain
\begin{align}
    \dot{S}_\alpha =& -\sum_\beta V_\alpha V_\beta \left(\frac{\qq_\alpha}{T_\alpha V_\alpha} + \frac{\qq_\beta}{T_\beta V_\beta}\right)\cdot W'_{\alpha\beta}\ee_{\alpha\beta} + \frac{1}{T_\alpha \tau}  \left( \frac{\qq_\alpha}{T_\alpha}\right)^2\\
    =& -V_\alpha D_\alpha^0 \left(\frac{\qq_\square}{T_\square V_\square}\right) +\frac{1}{T_\alpha \tau} (\tilde{G}_\alpha^0)^2.\nonumber
\end{align}
Heat flux \eqref{eq.wrelax} closer to the usual formula for heat flux \cite{jeong2003}, a difference being in the presence of particle volumes in the denominator. The overall entropy grows due to the last term, 
\begin{align}
    \frac{d}{dt}\sum_\alpha S_\alpha =& -\left\langle 1, D_\alpha^0\left(\frac{\qq_\square}{T_\square V_\square}\right)\right\rangle + \frac{1}{T_\alpha \tau} \sum_\alpha (\tilde{G}_\alpha^0)^2\nonumber\\
    =& \underbrace{\left\langle \tilde{G}_\alpha^0(1), \frac{\qq_\square}{T_\square V_\square}\right\rangle}_{=0} + \frac{1}{T_\alpha \tau} \sum_\alpha (\tilde{G}_\alpha^0)^2 \geq 0,
\end{align}
and the overall energy is conserved as well,
\begin{align}
    \frac{d E}{d t} =& \sum_\alpha T_\alpha \dot{S}_\alpha = -\left\langle T_\alpha, D_\alpha^0\left(\frac{\qq_\square}{T_\square V_\square}\right)\right\rangle + \frac{1}{\tau} \sum_\alpha \frac{T_\alpha}{T_\alpha \tau}  \left( \frac{\qq_\alpha}{T_\alpha}\right)^2\nonumber\\
    =& \underbrace{\left\langle \tilde{G}_\alpha^0(T_\square), \frac{\qq_\alpha}{T_\alpha V_\alpha}\right\rangle}_{=-\frac{1}{\tau}\sum_\alpha \frac{(\qq_\alpha)^2}{T^2_\alpha}}  + \frac{1}{\tau} \sum_\alpha \frac{T_\alpha}{\tau}  \left( \frac{\qq_\alpha}{T_\alpha}\right)^2=0.
\end{align}
and the overall energy is conserved as well. 
In summary, the model for hyperbolic heat conduction can be reduced to a model for Fourier heat conduction, where the relaxation parameter plays the role of heat conductivity. In the following Section, we derive an Eulerian approach to hyperbolic heat conduction model, which does not need any SPH operators.

\subsection{Hyperbolic heat conduction -- Eulerian approach}\label{sec.hyp.vor}
 The model for hyperbolic heat conduction in the preceding Section required the discrete SPH gradient in the definition of the Lagrangian conjugate entropy flux,  $\ww_\alpha$. However, there are several versions of SPH gradients which we might have chosen \cite{violeau}. Here, we proceed in an Eulerian way that does not need any beforehand specification the SPH gradient.

We add to the SPH state variables the Eulerian phonon momentum density attached to an SPH particle, $\ppi_{\alpha}$, which is related to the conjugate entropy flux by $\ppi_\alpha = \ww_\alpha S_\alpha$. The aim is to couple it with entropy via a modification of the Poisson bracket, where we take inspiration from the above knowledge of coupling particle momenta with particle entropy, in particular from the Poisson bracket \eqref{eq.PB.SPH-m.sa} with the mass-based volume. We consider the natural coupling between particle mass density and momentum as described in \eqref{eq.PB.SPH-m.sa} and we add a coupling between the auxiliary flux $\ppi_{\alpha}$ and entropy $s_{\alpha}$ as in the phonon Poisson bracket \cite{pkg}:
\begin{align}
\{F,G\}^{(\mathrm{SPH-m,\pi})} &= \{F,G\}^{(\mathrm{SPH})} \nonumber\\
    &+ \sum_\alpha\sum_\beta m_\beta W'_{\alpha\beta}e_{i\alpha\beta}\left(F_{\rho_\alpha}(G_{M_{\alpha i}}-G_{M_{\beta i}})
      -G_{\rho_\alpha}(F_{M_{\alpha i}}-F_{M_{\beta i}})\right)\nonumber\\
    &+ \sum_\alpha\sum_\beta S_\alpha\frac{m_\beta}{m_\alpha}  W'_{\alpha\beta}e_{i\alpha\beta}\left(F_{s_\alpha}(G_{\pi_{\alpha i}}-G_{\pi_{\beta i}}) -G_{s_\alpha}(F_{\pi_{\alpha i}}-F_{\pi_{\beta i}})\right). 
\end{align}
After the transformation to the total momentum $\bar{\MM}_\alpha = \MM_\alpha + \ppi_\alpha$, the bracket turns to
\begin{align}
    \{F,G\}^{(\mathrm{SPH-\bar{m},\pi})} &= \{F,G\}^{(\mathrm{SPH-mass})} \nonumber\\
    &+ \sum_\alpha\sum_\beta S_\alpha\frac{m_\beta}{m_\alpha}  W'_{\alpha\beta}e_{i\alpha\beta}\left(F_{s_\alpha}(G_{\pi_{\alpha i}}-G_{\pi_{\beta i}}) -G_{s_\alpha}(F_{\pi_{\alpha i}}-F_{\pi_{\beta i}})\right), 
\end{align}
that leads to an extension of evolution equations \eqref{eq.extSPH},
\begin{subequations}  \label{eq.hyp.eul}
    \begin{align}
        \dot{\xx}_\alpha &= E_{\bar{\MM}_\alpha}\\
        \dot{\bar{\MM}}_\alpha &= -E_{\xx_\alpha} - \sum_\beta m_\alpha m_\beta \left(\frac{E_{\rho_\alpha}}{m_\alpha} +\frac{E_{\rho_\beta}}{ m_\beta}+S_\alpha \frac{1}{m_\alpha^2} E_{s_\alpha}+S_\beta\frac{1}{m_\beta^2} E_{s_\beta}\right)W'_{\alpha\beta}\ee_{\alpha\beta}\\
        \dot{\rho}_\alpha &= \sum_\beta m_\beta W'_{\alpha\beta}\ee_{\alpha\beta}\cdot(E_{\bar{\MM}_{\alpha}}-E_{\bar{\MM}_\beta})\\
        \dot{s}_\alpha &= \frac{S_\alpha}{m_\alpha}\sum_\beta m_\beta W'_{\alpha\beta}\ee_{\alpha\beta}\cdot(E_{\bar{\MM}_{\alpha}}-E_{\bar{\MM}_\beta})
        +\sum_\beta S_\alpha \frac{m_\beta}{m_\alpha} \left(E_{\ppi_\alpha}-E_{\ppi_\beta}\right)\cdot W'_{\alpha\beta}\ee_{\alpha\beta}\\
        \dot{\ppi}_\alpha &= -\sum_\beta\left( S_\alpha \frac{m_\beta}{m_\alpha} E_{s_\alpha} + S_\beta \frac{m_\alpha}{m_\beta} E_{s_\beta}\right)W'_{\alpha\beta}\ee_{\alpha\beta}.
    \end{align}
\end{subequations} 
This set of evolution equations can be used to model a moving fluid with hyperbolic heat conduction.


\section{Conclusion}
In this paper, we compare five approaches towards SPH for nonbarotropic fluids, that is, for SPH with entropy. There are four ways to define the discrete particle volume (based on the mass, entropy,  a direct definition, or an implicit definition), and five ways to introduce the entropy (corresponding to the various particle volumes and to a mixed version). SPH with entropy conserves the total energy (dumping the kinetic energy into the internal energy). We formulate all five approaches as the sum of a symplectic part (reversible) and a conservative dissipative part (irreversible, raising the total entropy). 

The approach towards entropy using the mass-based particle volume, used, for instance, in the SDPD method \cite{espanol-revenga}, is perhaps the easiest to use because it does not change the usual SPH evolution equations, except for letting the pressure depend on the entropy. However, in cases with large inhomogeneities in entropy, the method does not provide sufficient detail.

The approach with the entropic volume improves the resolution of entropy density. However, it loses some details of the mass-density profile. Moreover, it is sensitive to negative particle entropies, which cause negative particle volumes.

The mixed approach combines the advantages of both the mass-based and the entropy-based particle volumes. On the other hand, the use of two particle volumes makes the approach ambiguous, and one should pay attention to the extra entropic terms in the evolution equations.

Finally, the implicit definition of the mass density and entropy density should be the most precise for interpolation. However, the price to pay is that the matrix inversion has to be carried out at each time step, which causes problems,  especially at the boundaries.

These various approaches towards entropy in SPH are illustrated in the adiabatic expansion of an ideal gas, nonadiabatic expansion, Rayleigh-Bénard convection without the Boussinesq approximation, and hyperbolic heat conduction.

In the future, we would like to apply these approaches towards entropy in numerical simulations of superfluid helium-4, where temperature waves (referred to as the second sound) are important when comparing with experiments \cite{bss,sciver,Mongiovi2018}.

\section*{Acknowledgement}
OK, MP, and VK were supported by the Czech Grant Agency (grant number 23-05736S). MP is a member of the Nečas Centre for Mathematical Modeling.

\appendix

\section{MaxEnt derivation of the SPH energy}\label{sec.MaxEnt}
The SPH energy \eqref{eq.SPH.ES} is an approximation of the exact continuum energy \eqref{eq.E.Lag}, and here we show that it can be obtained by means of the principle of maximum entropy (MaxEnt) \cite{jaynes}. The former depends on the SPH state variables, which are obtained by projection \eqref{eq.projection} from the state variables of the Lagrangian continuum mechanics. Therefore, the SPH energy expresses less information than the continuum energy. 

With a definition of energy on a more detailed level, one can identify the least biased estimate of the energy on the lower (less detailed) level by maximizing the detailed entropy constrained by the knowledge of the less detailed state variables \cite{shannon,jaynes}. To this end, we use the analogue of the MaxEnt principle, the principle of minimum energy, which is equivalent to MaxEnt for concave entropies \cite{callen}. The energy on the less detailed level of description (here SPH) follows from a double usage of the Legendre transform or, alternatively, from one Legendre transform and the use of the connection between the two levels (also called a projection) \cite{redext}. Thence, we seek minimizers of the
  \begin{equation}
    \label{eq:Phi}
    \phi (\xx,\MM;\xx_\alpha^\dagger,\MM_\alpha^\dagger)=-E^{Lagrangian}(\xx,\MM) + \langle(\pi_1(\xx,\MM),\pi_2(\xx,\MM)),(\xx_\alpha^\dagger,\MM_\alpha^\dagger) \rangle.
  \end{equation}
  The solution $\MM(\MM_\alpha^\dagger)$ to $$\frac{\delta \phi}{\delta M_i}=0$$
  reads $M_i=\rho_0 V_{0\alpha} \bar{\chi}_\alpha M_{\alpha,i}^\dagger$ while the solution $\xx(\xx_\alpha^\dagger,\MM_\alpha^\dagger)$ to  $$\frac{\delta \phi}{\delta x_i}=0$$
  is given by a condition
  $$\frac{\partial \epsilon}{\partial x_i} = \det\left( \frac{\partial \xx}{\partial \XX}\right) \bar{\chi}_\alpha x_{\alpha,i}^\dagger.$$
  The reduced conjugate energy is
  \begin{equation}
    \label{eq:E_SPH_conj}
    E^{SPH,\dagger}=\phi(\xx(\xx_\alpha^\dagger,\MM_\alpha^\dagger),\MM(\xx_\alpha^\dagger,\MM_\alpha^\dagger);\xx_\alpha^\dagger,\MM_\alpha^\dagger).
  \end{equation}
  For the particle-momentum contribution to the energy on the SPH level, we need to find the relation $\MM_\alpha(\MM_\alpha^\dagger)$ to identify the reduced energy which then follows from the momentum contribution to the conjugate SPH energy $\sum_\alpha=\frac{m_\alpha}{2} (\MM_\alpha^\dagger)^2$. Then, from the projection and $\MM(\MM_\alpha^\dagger)$, we get that the momentum contribution to the SPH energy is
  $\sum_\alpha\frac{1}{m_\alpha}(\MM_\alpha)^2$.
  The internal energy contribution to $\phi$ is approximated as follows
  \begin{multline}
    \label{eq:1}
 \int d\XX x_{\alpha,i}^\dagger\left(\frac{\xx(\XX)}{V_{0\alpha}}- \left(\frac{\partial \epsilon}{\partial x_i}\right)^{-1}\epsilon\right)= \frac{1}{V_{0\alpha}} \int_{V_{0\alpha}} d\XX x_{\alpha,i}^\dagger\xx(\XX)- V_\alpha \epsilon \\=  x_{\alpha,i}^\dagger x_{\alpha,i}- \frac{1}{V_{0\alpha}} \int_{V_{0\alpha}} d\XX V_\alpha \epsilon \approx x_{\alpha,i}^\dagger x_{\alpha,i}-  V_\alpha \epsilon_\alpha,
\end{multline}
where we introduced $\epsilon_\alpha(\rho_\alpha,s_\alpha)=\frac{1}{V_{0\alpha}} \int_{V_{0\alpha}} d\XX \epsilon(\rho(\XX),s(\XX)$. Therefore we have
\begin{equation}\label{eq.SPH.Ealpha}
    E^{SPH} = \sum_\alpha \frac{\MM^2_\alpha}{2 m_\alpha} +\sum_\alpha V_{\alpha}\epsilon_\alpha(\rho_\alpha, s_\alpha)
  \end{equation}
  as the Legendre transform of the conjugate SPH energy which removes $x_{\alpha,i}^\dagger x_{\alpha,i}$ and flips the sign. If the internal energy does not depend explicitly on the particle position, then this energy reduces to the SPH energy \eqref{eq.SPH.ES}. In other words, the SPH energy \eqref{eq.SPH.E} can be seen as the MaxEnt estimate of the energy subject the knowledge of the SPH state variables.

\section{Derivation of the Poisson bracket for the Entropic SPH}\label{sec.PB}
This Section contains details of the derivation of the Poisson brackets for the mass-volume, entropic-volume, mixed-volume, and implicit-volume approach.

\subsection{Mass-based volume Poisson bracket}\label{sec.PB.m}
Consider an arbitrary functional $F(\xx_\alpha, \MM_\alpha, \rho_\alpha, s_\alpha)$ of particle positions, momenta, densities, and entropy densities as defined in the main text, Eq. \eqref{eq:LagEulStateVars} with the particle volume $V_{\alpha}=V_{\alpha}^{m}$ as defined in Eq. \eqref{eq:Vm}. By plugging Equations \eqref{eq.projection} and \eqref{eq:LagEulStateVars} for particle mass and entropy density into bracket \eqref{eq.PB.Lag}, we obtain the Poisson bracket for the mass-based volume approach,
\begin{align}\label{eq.PB.SPH-m.sa}
\{F,G\}^{(\mathrm{SPH-mass})}&= \{F,G\}^{(\mathrm{SPH})} \nonumber\\
    &+ \sum_\alpha\sum_\beta m_\beta W'_{\alpha\beta}e_{i\alpha\beta}\left(F_{\rho_\alpha}(G_{M_{\alpha i}}-G_{M_{\beta i}})
    -G_{\rho_\alpha}(F_{M_{\alpha i}}-F_{M_{\beta i}})\right)\nonumber\\
    &+ \sum_\alpha\sum_\beta S_\alpha \frac{m_\beta}{m_\alpha} W'_{\alpha\beta}e_{i\alpha\beta}\left(F_{s_\alpha}(G_{M_{\alpha i}}-G_{M_{\beta i}})
    -G_{s_\alpha}(F_{M_{\alpha i}}-F_{M_{\beta i}})\right).
\end{align}
This Poisson bracket leads to the Hamiltonian equations \eqref{eq.extSPH}.

\subsection{Entropic-based volume Poisson bracket}\label{eq.PB.s}
Consider an arbitrary functional $F(\xx_\alpha, \MM_\alpha, \rho_\alpha, s_\alpha)$ of particle positions, momenta, densities, and entropy densities as defined in the main text, Eq. \eqref{eq:LagEulStateVars} with the particle volume $V_{\alpha}=V_{\alpha}^{s}$ as defined in Eq. \eqref{eq:Vs}. By plugging Equations \eqref{eq.projection} and \eqref{eq:LagEulStateVars} for particle mass and entropy density into bracket \eqref{eq.PB.Lag}, we obtain the Poisson bracket for the entropic-based volume approach,
\begin{align}\label{eq.PB.SPH-s.rhoa}
\{F,G\}^{(\mathrm{SPH-entropic})}&= \{F,G\}^{(\mathrm{SPH})} \nonumber\\
    &+ \sum_\alpha\sum_\beta S_\beta W'_{\alpha\beta}e_{i\alpha\beta}\left(F_{s_\alpha}(G_{M_{\alpha i}}-G_{M_{\beta i}})
    -G_{s_\alpha}(F_{M_{\alpha i}}-F_{M_{\beta i}})\right)\nonumber\\
    &+ \sum_\alpha\sum_\beta m_\alpha \frac{S_\beta}{S_\alpha} W'_{\alpha\beta}e_{i\alpha\beta}\left(F_{\rho_\alpha}(G_{M_{\alpha i}}-G_{M_{\beta i}})
    -G_{\rho_\alpha}(F_{M_{\alpha i}}-F_{M_{\beta i}})\right).
\end{align}

 \subsection{Mixed-volume Poisson bracket}\label{sec.PB.x}
 Consider now an arbitrary functional $F(\xx_\alpha, \MM_\alpha, \rho_\alpha, s_\alpha)$ of positions, momenta, densities, and entropy densities with the latter defined by Equations \eqref{eq.mixed}. Before plugging such functionals into the Lagrangian Poisson bracket \eqref{eq.PB.Lag}, let us compute the derivatives of the functional. The derivative with respect to the position becomes
 \begin{align}
     \frac{\delta F}{\delta x^i(\XX)}  &= \sum_\alpha \frac{\partial F}{\partial x^i_\alpha} \chibar_\alpha(\XX) 
     +\sum_\alpha\sum_\gamma \frac{\partial F}{\partial \rho_\alpha}\frac{\partial \rho_\alpha}{\partial x_\gamma^j}\frac{\delta x^j_\gamma}{\delta x^i(\XX)}\nonumber\\
     &+\sum_\alpha\sum_\gamma \frac{\partial F}{\partial s_\alpha}\frac{\partial s_\alpha}{\partial x_\gamma^j}\frac{\delta x^j_\gamma}{\delta x^i(\XX)}\nonumber\\
     &= \sum_\alpha \frac{\partial F}{\partial x^i_\alpha} \chibar_\alpha(\XX) 
     +\sum_\alpha \frac{\partial F}{\partial \rho_\alpha} \sum_\beta\sum_\gamma m_\beta W'_{\alpha\beta}e_{i\alpha\beta}(\delta_{\alpha\gamma}-\delta_{\beta\gamma})\chibar_{\gamma}(\XX)\nonumber\\
     &+\sum_\alpha \frac{\partial F}{\partial s_\alpha} \sum_\beta\sum_\gamma S_\beta W'_{\alpha\beta}e_{i\alpha\beta}(\delta_{\alpha\gamma}-\delta_{\beta\gamma})\chibar_{\gamma}(\XX).
 \end{align}

 Plugging such two functionals into Poisson bracket \eqref{eq.PB.Lag} then leads to 
 \begin{align}\label{eq.PB.x}
     \{F,G\}^{(\mathrm{SPH-mixed})} &= \{F,G\}^{(SPH)} \nonumber\\
     &+ \sum_\alpha\sum_\beta\sum_\gamma\sum_\delta F_{\rho_\alpha} m_\beta W'_{\alpha\beta}e_{i\alpha\beta}(\delta_{\alpha\gamma}-\delta_{\beta\gamma})G_{M_{\delta i}}\int d\XX V_{0\delta}\chibar_\gamma(\XX)\chibar_\delta(\XX) \nonumber\\
     &- \stackrel{F\leftrightarrow G}{\dots}\nonumber\\
     &+ \sum_\alpha\sum_\beta\sum_\gamma\sum_\delta F_{s_\alpha} S_{\beta} W'_{\alpha\beta}e_{i\alpha\beta}(\delta_{\alpha\gamma}-\delta_{\beta\gamma})G_{M_{\delta i}}\int d\XX V_{0\delta}\chibar_\gamma(\XX)\chibar_\delta(\XX) \nonumber\\
     &- \stackrel{F\leftrightarrow G}{\dots}\nonumber\\
     &= \{F,G\}^{(SPH)} \nonumber\\
     &+ \sum_\alpha\sum_\beta\sum_\gamma F_{\rho_\alpha}m_\beta W'_{\alpha\beta}e_{i\alpha\beta}(\delta_{\alpha\gamma}-\delta_{\beta\gamma})G_{M_{\gamma i}}- \stackrel{F\leftrightarrow G}{\dots}\nonumber\\
     &+ \sum_\alpha\sum_\beta\sum_\gamma F_{s_\alpha}S_{\beta} W'_{\alpha\beta}e_{i\alpha\beta}(\delta_{\alpha\gamma}-\delta_{\beta\gamma})G_{M_{\gamma i}}- \stackrel{F\leftrightarrow G}{\dots}\nonumber\\
     &= \{F,G\}^{(SPH)} \nonumber\\
     &+ \sum_\alpha\sum_\beta m_\beta W'_{\alpha\beta}e_{i\alpha\beta}\left(F_{\rho_\alpha}(G_{M_{\alpha i}}-G_{M_{\beta i}}
     -G_{\rho_\alpha}(F_{M_{\alpha i}}-F_{M_{\beta i}})\right)\nonumber\\
     &+ \sum_\alpha\sum_\beta S_{\beta} W'_{\alpha\beta}e_{i\alpha\beta}\left(F_{s_\alpha}(G_{M_{\alpha i}}-G_{M_{\beta i}})
     -G_{s_\alpha}(F_{M_{\alpha i}}-F_{M_{\beta i}})\right),
 \end{align}
 which is the Poisson bracket for the approach towards the entropic SPH with the mixed volume. This Poisson bracket then leads to reversible volution equations \eqref{eq.extSPH-x}.

\subsection{Implicit-volume Poisson bracket}\label{sec.PB.I}
Consider now an arbitrary functional $F(\xx_\alpha, \MM_\alpha, \rho_\alpha, s_\alpha)$ of the particle positions, momenta, densities, and entropy densities as defined in the main text, Eq. \eqref{eq:LagEulStateVars} with the particle volume $V_{\alpha}=\tilde{V}_{\alpha}^{I}$ as defined in Eq. \eqref{eq:VI}. Before plugging such functionals into the Lagrangian Poisson bracket \eqref{eq.PB.Lag}, let us compute the derivatives of the functional. The derivative with respect to the position becomes
\begin{align}
  \frac{\delta F}{\delta x^i(\XX)}  &= \sum_\alpha \frac{\partial F}{\partial x^i_\alpha} \chibar_\alpha(\XX) +\sum_\alpha \frac{\partial F}{\partial \rho_\alpha}\frac{\delta \rho_\alpha}{\delta x^i(\XX)}+\sum_\alpha \frac{\partial F}{\partial s_\alpha}\frac{\delta s_\alpha}{\delta x^i(\XX)}\nonumber\\
                                    &= \sum_\alpha \frac{\partial F}{\partial x^i_\alpha} \chibar_\alpha(\XX)  +\sum_\alpha \frac{\partial F}{\partial \rho_\alpha}  \sum_\beta\sum_\gamma \frac{m_\beta}{\rho_\beta} \frac{\rho_\alpha^2}{m_\alpha} W_{\alpha\gamma}^{-1} W_{\gamma\beta}' e_{i \delta\gamma} (\chibar_\gamma-\chibar_\beta)\nonumber\\
    &+\sum_\alpha \frac{\partial F}{\partial s_\alpha} \sum_\beta\sum_\gamma \frac{S_\beta}{s_\beta} \frac{s_\alpha^2}{S_\alpha} W_{\alpha\gamma}^{-1} W_{\gamma\beta}' e_{i \delta\gamma} (\chibar_\gamma-\chibar_\beta),
\end{align}
where the partial derivatives $\frac{\partial \rho_\alpha}{\delta x^i(\XX)}$ follow from the implicit definition of volume $1=\sum_{\beta} \tilde{V}_{\alpha}^{I} W_{\alpha\beta}$ noting that $\tilde{V}_{\alpha}^{I}=\frac{m_{\alpha}}{\rho_{\alpha}}=\frac{S_{\alpha}}{s_{\alpha}}$ for all $\alpha$. Hence
\begin{equation*}
  0=-\frac{m_\delta}{\rho_\delta^2}\frac{\delta \rho_\delta}{\delta x^i(\XX)}+\sum_\alpha\sum_\beta\sum_\gamma \frac{m_\beta}{\rho_\beta}W^{-1}_{\delta\alpha} W_{\alpha\beta}' e_{i\alpha\beta}(\delta_{\alpha\gamma}-\delta_{\beta\gamma})\chibar_\gamma(\XX)
\end{equation*}
and thus
\begin{equation*}
  \frac{\delta \rho_\alpha}{\delta x^i(\XX)} = \sum_\beta\sum_\gamma \frac{m_\beta}{\rho_\beta} \frac{\rho_\alpha^2}{m_\alpha} W_{\alpha\gamma}^{-1} W_{\gamma\beta}' e_{i \delta\gamma} (\chibar_\gamma-\chibar_\beta).
\end{equation*}
Similarly, one obtains the used expression for $ \frac{\delta s_\alpha}{\delta x^i(\XX)}$ in the implicit-based volume case. Note that we again assumed that Lagrangian particle mass and entropy are independent of the particle positions (although $S_{\alpha}$ can change in time due to irreversible effects).

Plugging such two functionals into Poisson bracket \eqref{eq.PB.Lag} then leads to the Poisson bracket for the implicit-based volume approach towards the entropic SPH
\begin{align}\label{eq.PB.SPH-I.rhoa.sa}
    \{F,G\}^{(\mathrm{SPH-implicit})} &= \{F,G\}^{(SPH)} \\
                      &+ \sum_\alpha\sum_\beta\sum_\delta F_{\rho_\alpha} \frac{m_\beta \rho_\alpha^2}{\rho_\beta m_\alpha} W_{\alpha\delta}^{-1} W_{\delta\beta}' e_{i\delta\beta}(G_{M_i^\delta}-G_{M_i^\beta}) + \sum_\alpha\sum_\beta\sum_\delta F_{s_\alpha} \frac{S_\beta s_\alpha^2}{s_\beta S_\alpha}W_{\alpha\delta}^{-1} W_{\delta\beta}' e_{i\delta\beta}(G_{M_i^\delta}-G_{M_i^\beta})\nonumber\\
  &-\sum_\alpha\sum_\beta\sum_\delta G_{\rho_\alpha} \frac{m_\beta \rho_\alpha^2}{\rho_\beta m_\alpha} W_{\alpha\delta}^{-1} W_{\delta\beta}' e_{i\delta\beta}(F_{M_i^\delta}-F_{M_i^\beta}) - \sum_\alpha\sum_\beta\sum_\delta G_{s_\alpha} \frac{S_\beta s_\alpha^2}{s_\beta S_\alpha}W_{\alpha\delta}^{-1} W_{\delta\beta}' e_{i\delta\beta}(F_{M_i^\delta}-F_{M_i^\beta}).\nonumber
\end{align}

\section{Thermodynamics}\label{sec.thermo}
This Section recalls standard relations on the ideas gas model and the model of stiffened gas \cite{callen,ader-vis}.

\subsection{Ideal gas}
The fundamental thermodynamic relation for an ideal gas reads
\begin{equation}
\epsilon = \frac{\rho^\gamma}{\gamma-1}e^{\frac{s}{\rho c_V}}
\end{equation}
where $\epsilon$ is the volumetric energy density, $s$ is the volumetric entropy density, and $\gamma=\frac{c_P}{c_V}$. From this equation it follows that
\begin{subequations}
\begin{align}
p =& \rho^\gamma e^{\frac{s}{c_V \rho}} = (\gamma-1)c_V \rho T\\
T =& \frac{\rho^{\gamma-1}}{c_V(\gamma-1)} e^{\frac{s}{c_V\rho}}\\
\mu =& \frac{\gamma \rho^{\gamma-1}}{\gamma-1} e^{\frac{s}{c_V\rho}} - \frac{\rho^{\gamma-2}}{\gamma-1} e^{\frac{s}{c_V\rho}} \frac{s}{c_V}
\end{align}
and $\epsilon=\rho c_V T$.
\end{subequations}

\subsection{Stiffened gas}\label{sec.stiffened}
The fundamental thermodynamic relation of a stiffened generalizes that of the ideal gas to
\begin{equation}
\epsilon = \rho\left(\frac{c_0^2}{\gamma(\gamma-1)} \left(\frac{\rho}{\rho_0}\right)^{\gamma-1} e^{\frac{s}{c_V\rho}} + \frac{\rho_0 c_0^2 -\gamma p_0}{\gamma \rho}\right)
\end{equation}
where $c_0$ is a reference speed of sound, $\rho_0$ is a reference density, and $p_0$ is a reference pressure.
This leads to 
\begin{subequations}
\begin{align}
p =& (\gamma-1)\epsilon -(\rho_0 c_0^2 -\gamma p_0)\\
T =& \frac{c_0^2}{c_V \gamma(\gamma-1)}\left(\frac{\rho}{\rho_0}\right)^{\gamma-1}e^{\frac{s}{c_V \rho}}\\
\mu =& \frac{c_0^2}{\gamma-1}\left(\frac{\rho}{\rho_0}\right)^{\gamma-1} e^{\frac{s}{\rho c_V}} -\rho \frac{c_0^2}{\gamma(\gamma-1)}\left(\frac{\rho}{\rho_0}\right)^{\gamma-1}\frac{s}{\rho^2 c_V}
\end{align}
\end{subequations}
In particular, if we set $\rho=\rho_0$ and $p=p_0$, we get $s = 0$ and that the speed of sound becomes $c_0$. Moreover, the equation of state reads 
\begin{equation}
    p = (\gamma-1)\rho c_V T - \frac{\rho_0 c_0^2 - \gamma p_0}{\gamma}.
\end{equation}

\end{document}